\documentclass[useAMS,usenatbib,usegraphicx]{mn2e}
\usepackage{multirow}
\usepackage{amsmath}
\title[Radio-loudness and central surface brightness profiles]{The connection between radio-loudness and central surface brightness profiles in optically-selected low-luminosity active galaxies}
\author[A. J. Richings, P. Uttley and E. K\"{o}rding]{A. J. Richings$^{1}$, P. Uttley$^{1}$ and E. K\"{o}rding$^{2}$\\
$^{1}$School of Physics and Astronomy, University of Southampton, Southampton, S017 1BJ, UK\\
$^{2}$Radboud Universiteit Nijmegen, Dept. Astronomy, IMAPP, The Netherlands}
\begin{document}

\date{\today}

\pagerange{\pageref{firstpage}--\pageref{lastpage}} \pubyear{2011}

\maketitle

\label{firstpage}

\begin{abstract}
Recent results indicate a correlation between nuclear radio-loudness of active galaxies and their central stellar surface-brightness profiles, in that `core' galaxies (with inner logarithmic slope $\gamma\leq 0.3$) are significantly more radio loud than `power-law' galaxies ($\gamma\geq 0.5$).  This connection, which indicates possible links between radio-loudness and galaxy formation history (e.g. through black hole spin) has so far only been confirmed for a radio-selected sample of galaxies. Furthermore, it has since been shown that the Nuker law, which was used to parameterise the brightness profiles in these studies, gives a poor description of the brightness profile, with its parameters varying systematically with the radial fitted extent of the profile. Here, we present an analysis of the central surface brightness profiles of the active galaxies of Hubble Type $T\leq3$, that were identified by the optically-selected Palomar spectroscopic survey of nearby galaxies. We fit the brightness profiles using S\'{e}rsic, Core-S\'{e}rsic and, where necessary, Double-S\'{e}rsic models, which we fit to the semi-major axis brightness profiles extracted from high resolution images of the galaxies from the Hubble Space Telescope ({\it HST}). We use these fits to classify the galaxies as `Core', `S\'{e}rsic' or `Double-S\'{e}rsic'. We compare the properties of the Active Galactic Nuclei (AGNs) and their host galaxies with this classification, and we recover the already established trend for Core galaxies to be more luminous and contain a higher-mass supermassive black hole. Defining the radio-loudness of an AGN as the ratio of the nuclear radio luminosity to [O{\rm {\sc iii}}] line luminosity, which allows us to include most of the AGN in our sample and prevents a bias against dim nuclei that are harder to extract from the brightness profiles, we find that AGN hosted in Core galaxies are generally more radio-loud than those hosted in S\'{e}rsic galaxies, although there is a large overlap between the two subsamples.  The correlation between radio-loudness and brightness profile can partly be explained by a correlation between radio-loudness and black hole mass. Additionally, there is a significant (99 per cent confidence) partial correlation between radio-loudness and the Core/S\'{e}rsic classification of the host galaxy, which lends support to the previous results based on the radio-selected sample, although it is possible that this partial correlation arises because AGN in core galaxies tend to have a lower accretion rate as well as a higher central black hole mass. 
\end{abstract}


\section{Introduction}
Before the {\it Hubble Space Telescope} ({\it HST}) was launched, observations of the central regions of galaxies were limited by atmospheric seeing, which prevented resolutions better than $\sim1\arcsec$ from being achieved. This made it difficult to determine how the brightness profiles of these galaxies behaved at small radii. However, high resolution images taken with the {\it HST} \citep{crane, ferrarese, fb} demonstrated that the surface brightness profiles of elliptical galaxies and disk galaxy bulges continue to increase up to the resolution limit, confirming the results of ground-based observations \citep{kormendy, lauer85} that concluded that the profiles of most elliptical galaxies cannot be described by models with flat, isothermal cores such as described by \citet{king}.

Early studies using {\it HST} suggested that the central brightness profiles can be divided into two distinct classes --- in ``power-law'' galaxies, the logarithmic slope of the brightness profile remained steep towards the resolution limit of {\it HST}, while in ``core'' galaxies the slope flattened significantly. To describe these brightness profiles in the central regions, \citet{la} introduced a form of broken power law, the so-called ``Nuker-law'', with the logarithmic slope in the inner region, $\gamma$, being flatter than in the outer region. These classes then correspond to a dichotomy in the values of $\gamma$, with power-law galaxies having $\gamma\geq0.5$ and core galaxies having $\gamma\leq0.3$. \citet{fa} demonstrated that there were no galaxies in the range $0.3<\gamma<0.5$, although subsequent studies \citep[e.g.][]{rb} have found a small number of such ``intermediate'' galaxies.

This dichotomy in brightness profiles has been linked to global galaxy properties, for example core galaxies tend to be very luminous, slowly rotating galaxies with boxy isophotes, while power-law galaxies tend to be less luminous, rapidly rotating galaxies with disky isophotes (\citealt{ja}; \citealt{fa}). For galaxies that contain active galactic nuclei (AGNs) some of these properties of the host galaxy also correlate with properties of the AGN, for example the mass of the central supermassive black hole (SMBH) correlates strongly with the velocity dispersion of the host galaxy \citep{ferrarese00,gebhardt}, suggesting that the evolution of the host galaxy and AGN are interdependent. Investigating how the brightness profiles of the central regions of these host galaxies are related to the properties of the AGN is important in understanding how they influence one another and how they form and evolve together, for example \citet{fa} propose a model whereby a merger between two galaxies can produce a core galaxy as their black holes sink to the centre, ejecting stars in the central region and scouring out a core, while \citet{va} suggests that the cusp slopes of core and power-law galaxies can be reproduced by adiabatic growth of the black hole in an isothermal core.

More recent studies \citep[e.g.][]{graham03,ferrarese06a} have shown that the Nuker model does not give an accurate parameterization of galaxy surface brightness profiles. It was only intended to model the central regions and so cannot describe the brightness profiles at larger radii, where the outer power law of the Nuker model fails to properly fit the curvature of the brightness profiles in most galaxies, and \citet{graham03} demonstrate that the parameters of the Nuker model are sensitive to the radial extent of the profile to which it is fitted. However, the brightness profiles of ``power-law'' galaxies are well fitted by either a S\'{e}rsic law over their entire radial extent, or by a S\'{e}rsic law modified with the addition of a second S\'{e}rsic component, the latter meant to represent an inner compact stellar nucleus (generally below a few tenths of an arcsec). Meanwhile, ``core'' galaxies show a central deficit with respect to their outer S\'{e}rsic profile \citep{trujillo,ferrarese06b}. This provides an alternative classification to the core/power law dichotomy based on inner logarithmic slope, in which galaxies can be separated into those that are well fitted by a S\'{e}rsic law at all radii, those that require an additional inner S\'{e}rsic component to account for an inner compact stellar nucleus and those that show a central deficit from their outer S\'{e}rsic profile. The profiles of galaxies that show a central deficit can be parameterized by a ``Core-S\'{e}rsic'' model \citep[introduced by][]{graham03} that combines a S\'{e}rsic profile at large radii with a power law in the central regions \citep[see e.g.][]{graham04, cote07}.

Recently, Capetti \& Balmaverde published a series of three papers (\citealt{cb} \& 2006 and \citealt{cc}) in which they investigate the core/power-law dichotomy, based on the Nuker law, in two radio-selected samples of galaxies that are likely to host an AGN, requiring a minimum radio flux of $\sim1$~mJy. Using new Nuker fits to the data as well as fits from the literature, they showed that radio-loud AGNs (defined using the ratio of 5~GHz to B-band luminosity densities) are hosted by core galaxies, while radio-quiet AGNs are hosted by power-law galaxies. \citet{cb07} also showed that early-type galaxies hosting Seyfert nuclei show power-law type central brightness profiles.  This correlation of AGN radio-loudness with brightness profile offers important clues to the origin of the radio-quiet/radio-loud dichotomy in AGN, as well as the link between the central black hole and the host galaxy.  For example, if the merger history governs whether a galaxy is core or power-law, we might speculate that core galaxies host rapidly spinning black holes resulting from a major merger, and that radio-loudness is linked to spin.  Alternatively, the different stellar density profiles in core and power-law galaxies may be linked to different environments which may enhance or inhibit jet production, e.g. lower or higher gas densities respectively, and/or different types of hot or cold accretion (e.g. \citealt{best05}).  Regardless of its physical interpretation, the radio-loudness/brightness-profile connection may shed useful light on the connection between radio-loud AGN and giant elliptical hosts, since core type galaxies are almost all giant ellipticals.

The implications of the radio-loudness/brightness-profile connection in AGN make it worthy of further study.  In particular, it is important to identify whether the link persists in an optically selected sample of AGN.  This is because the Capetti \& Balmaverde sample was radio-selected, so that it is still possible that there exists a population of AGN which inhabit core galaxies but are sufficiently radio-quiet that they were not selected.  Clearly it is technically very difficult to measure central brightness profiles for optically luminous AGN, but in any case the Capetti \& Balmaverde sample is dominated by AGN with relatively low optical luminosities, i.e. predominantly optically classified as Low Ionisation Nuclear Emission Regions (LINERs) or other Low Luminosity Active Galactic Nuclei (LLAGNs).

In order to search for such AGN in nearby galaxies, \citet{ha} conducted an optical spectroscopic survey of the nuclei of bright, nearby galaxies in the Northern hemisphere, the so-called Palomar spectroscopic survey of nearby galaxies. Their survey included 197 galaxies that are confirmed to contain an AGN, as measured by their emission line ratios. This gives us an optically selected sample of low-luminosity AGNs, in contrast to Capetti \& Balmaverde's radio-selected sample. In this paper, we use available {\it HST} images to analyse the brightness profiles of the Palomar survey active galaxies, using the S\'{e}rsic/Core-S\'{e}rsic classification scheme rather than the core/power-law classification obtained from the Nuker model. We then use these results to investigate how the radio properties of these AGNs are related to their S\'{e}rsic/Core-S\'{e}rsic classification, which allows us to show whether Capetti \& Balmaverde's results are reproduced in a relatively complete sample of optically-selected AGN using this alternative classification scheme.

\section{Sample and data reduction}

\subsection{The Palomar spectroscopic survey of nearby galaxy nuclei}
\label{sample}
To reliably detect the presence of a galaxy core it must be nearby so that a core could easily be resolved by the {\it HST}. The Palomar spectroscopic survey of nearby galaxy nuclei \citep{ha} uses a flux-limited sample of 488 bright Northern galaxies, with $B_T\leq12.5$ mag and declinations $\delta>0\degr$, and naturally satisfies our requirement for nearby galaxies. Spectroscopic observations of these galaxies show that they include 197 AGNs. \citet{na} conducted a high resolution radio survey of all of these AGNs, which will enable us to compare their radio properties to their Core/S\'{e}rsic classification. For spiral galaxies it is the brightness profile of the bulge component that determines this classification, however late-type spiral galaxies have relatively small bulges compared to the disk component, which will make it more difficult to classify them. Therefore we take as our sample the 150 AGNs in \citet{na} for which the host galaxy is of Hubble type $T\leq3$.

\citet{hb} list various parameters for the galaxies in the Palomar sample, such as distances, Hubble types, AGN classification and bulge and total B-band absolute magnitudes, which we use in this paper. However, where galaxies have more recent distance estimates listed in \citet{tonry01} and \citet{paturel02}, we have used these values, and updated the absolute magnitudes and luminosities accordingly.  The [O{\rm {\sc iii}}] and H$\beta$ narrow line luminosities for the AGN were obtained using the combination of unreddened line ratio data and H$\alpha$ fluxes in Table 2 of \citet{hb}.  We dereddened these fluxes by applying the extinction curve of \citet{cardelli89}, assuming the Balmer decrement to be 3.1.

We also estimated black hole masses for the AGN in our subset of the Palomar sample using the central velocity dispersions measured by \citet{ho09}. We converted velocity dispersion to black hole mass using the relation: $\log(M_{\rm BH}/{\rm M_{\odot}})=8.13+4.02\log(\sigma/200 {\rm km s^{-1}})$ \citep{ta}.  For consistency this method was used for all of the galaxies, including those for which direct kinematic measurements were also available.

\subsection{HST data reduction}
{\it HST} images of the galaxies in our sample were obtained from the public archive. Previous studies \citep[e.g.][]{ra} have demonstrated that many early-type galaxies contain dust structures that may affect the brightness profiles. It is especially important to deal with the effects of dust in the AGN sample we study here, since dust structures are almost ubiquitous in LLAGN \citep{gonzalez-delgado} and are probably more common in LLAGN than in normal galaxies \citep{martini}. Therefore to minimise the effects of dust we preferred to use images taken in the near-infrared using NICMOS, with the F160W filter. However, for many of the galaxies in our sample NICMOS images were unavailable, so for these we used images taken in the optical using WFPC2 or ACS. Only broadband filters were used, and preference was given to filters at longer wavelengths --- any shorter than F555W were rejected. Using different filters might affect our results if the brightness profile depends on the wavelengths at which the galaxy is observed, however \citet{rb} showed that the classification of a galaxy as core or power-law, based on Nuker fits, is independent of the wavelength that is used. We were able to obtain suitable images for 110 galaxies in our sample; the basic properties of these galaxies are presented in table~\ref{basicprop}. As described in Section~\ref{sample} the data presented in this table were obtained from \citet{hb}, although data was only available for 107 of the 110 Palomar survey galaxies for which we were able to obtain {\it HST} images.  

\begin{table*}
\begin{minipage}{168mm}
\caption{Basic properties of the galaxies in our sample for which {\it HST} images were obtained.}
\begin{tabular}{lccccccccll}
\hline
Galaxy & T & D & $M_{B, total}$ & $M_{B, bulge}$ & $\log\left(\frac{M_{BH}}{M_{\sun}}\right)$ & $\log L_{[OIII]}$ & $\log L_{H\beta}$ & $\log L_{\nu, radio}$ & AGN & Class \\
\hline
IC0356	&	2	&	18.1	&	-21.12	&	-19.89	&	7.70	&	39.27	&	38.94	&	$<$36.29	&	T2	&	S\'{e}rsic?	\\
NGC0315	&	-4	&	65.8	&	-22.22	&	-22.22	&	8.86	&	39.44	&	39.61	&	40.09	&	L1.9	&	Core?	\\
NGC0404	&	-3	&	3.3	&	-16.65	&	-16.18	&	5.32	&	37.69	&	37.6	&	$<$34.92	&	L2	&	Double	\\
NGC0474	&	-2	&	32.5	&	-20.59	&	-19.98	&	7.78	&	38.7	&	38.34	&	$<$36.98	&	L2::	&	S\'{e}rsic	\\
NGC0488	&	3	&	29.3	&	-21.43	&	-19.89	&	8.03	&	38.69	&	38.74	&	$<$36.71	&	T2::	&	S\'{e}rsic	\\
NGC0524	&	-1	&	24	&	-20.73	&	-20	&	8.54	&	$<$37.55	&	37.87	&	36.71	&	T2:	&	Core	\\
NGC0660	&	1	&	11.8	&	-18.92	&	-17.9	&	7.35	&	40.28	&	39.96	&	$<$35.88	&	T2/H:	&	S\'{e}rsic?	\\
NGC2273	&	0.5	&	28.4	&	-20.25	&	-19.32	&	7.61	&	41.23	&	40.5	&	37.39	&	S2	&	S\'{e}rsic?	\\
NGC2681	&	0	&	17.2	&	-20.3	&	-19.44	&	7.07	&	39.23	&	39.49	&	$<$36.40	&	L1.9	&	S\'{e}rsic?	\\
NGC2683	&	3	&	7.7	&	-19.82	&	-18.28	&	7.38	&	37.72	&	37.27	&	$<$35.51	&	L2/S2	&	S\'{e}rsic?	\\
NGC2685	&	-1	&	16.2	&	-19.23	&	-18.5	&	6.81	&	39.23	&	38.77	&	$<$36.15	&	S2/T2:	&	S\'{e}rsic	\\
NGC2768	&	-5	&	22.4	&	-21.05	&	-21.05	&	7.96	&	38.43	&	38.24	&	37.39	&	L2	&	Double?	\\
NGC2787	&	-1	&	7.5	&	-17.76	&	-17.03	&	8.15	&	37.74	&	38.02	&	36.37	&	L1.9	&	Double?	\\
NGC2832	&	-4	&		&		&		&	9.03	&		&		&		&		&	Core	\\
NGC2841	&	3	&	12	&	-20.82	&	-19.28	&	8.31	&	38.57	&	38.31	&	36.26	&	L2	&	Core	\\
NGC2859	&	-1	&	25.4	&	-20.38	&	-19.65	&	8.02	&	38.53	&	38.1	&	$<$36.59	&	T2:	&	Double?	\\
NGC2985	&	2	&	22.4	&	-20.77	&	-19.54	&	7.52	&	37.85	&	38.07	&	$<$36.48	&	T1.9	&	S\'{e}rsic	\\
NGC3169	&	1	&	19.7	&	-20.51	&	-19.49	&	8.03	&	39.46	&	39.03	&	37.2	&	L2	&	Double	\\
NGC3190	&	1	&	22.4	&	-20.31	&	-19.29	&	8.02	&	38.67	&	38.35	&	36.38	&	L2	&	S\'{e}rsic?	\\
NGC3193	&	-5	&	34	&	-20.93	&	-20.93	&	8.08	&	38.46	&	38.06	&	$<$37.02	&	L2:	&	Core	\\
NGC3227	&	1	&	16.5	&	-19.91	&	-18.89	&	7.46	&	40.3	&	40.41	&	36.88	&	S1.5	&	S\'{e}rsic	\\
NGC3245	&	-2	&	20.9	&	-19.95	&	-19.34	&	8.21	&	39.27	&	39.49	&	$<$36.42	&	T2:	&	S\'{e}rsic	\\
NGC3368	&	2	&	10.4	&	-20.28	&	-19.05	&	7.35	&	38.85	&	38.63	&	$<$35.81	&	L2	&	S\'{e}rsic?	\\
NGC3379	&	-5	&	10.6	&	-19.94	&	-19.94	&	8.19	&	37.96	&	37.69	&	$<$35.83	&	L2/T2::	&	Core	\\
NGC3414	&	-2	&	25.2	&	-20.15	&	-19.54	&	8.42	&	39.14	&	38.8	&	36.96	&	L2	&	S\'{e}rsic	\\
NGC3489	&	-1	&	12.1	&	-19.26	&	-18.53	&	7.12	&	38.73	&	38.27	&	$<$35.94	&	T2/S2	&	S\'{e}rsic?	\\
NGC3507	&	3	&	19.8	&	-19.86	&	-18.32	&	6.65	&	39.09	&	39.11	&	$<$36.55	&	L2	&	S\'{e}rsic?	\\
NGC3516	&	-2	&	37.9	&	-20.75	&	-20.14	&	7.96	&	40.44	&	41	&	37.05	&	S1.2	&	S\'{e}rsic	\\
NGC3607	&	-2	&	22.8	&	-21	&	-20.39	&	8.39	&	39.48	&	39.16	&	36.64	&	L2	&	Core	\\
NGC3608	&	-5	&	22.9	&	-20.11	&	-20.11	&	8.06	&	38.24	&	37.77	&	$<$36.67	&	L2/S2:	&	Core	\\
NGC3623	&	1	&	7.3	&	-19.67	&	-18.65	&	7.48	&	37.65	&	37.28	&	$<$35.46	&	L2:	&	S\'{e}rsic?	\\
NGC3626	&	-1	&	20	&	-19.89	&	-19.16	&	7.58	&	39.19	&	39.15	&	$<$36.56	&	L2:	&	S\'{e}rsic?	\\
NGC3627	&	3	&	9.1	&	-20.66	&	-19.12	&	7.30	&	39.37	&	38.95	&	36.16	&	T2/S2	&	S\'{e}rsic?	\\
NGC3628	&	3	&	7.7	&	-20.12	&	-18.58	&	6.47	&	37.04	&	36.82	&	35.89	&	T2	&	S\'{e}rsic?	\\
NGC3675	&	3	&	12.8	&	-20.02	&	-18.48	&	7.05	&	37.83	&	37.73	&	$<$35.99	&	T2	&	Double?	\\
NGC3705	&	2	&	17	&	-19.9	&	-18.67	&	7.18	&	38.71	&	38.7	&	$<$36.32	&	T2	&	Double?	\\
NGC3718	&	1	&	17	&	-19.96	&	-18.94	&	7.72	&	38.14	&	38.17	&	37.27	&	L1.9	&	S\'{e}rsic	\\
NGC3898	&	2	&	21.9	&	-20.38	&	-19.15	&	8.19	&	38.74	&	38.42	&	$<$36.46	&	T2	&	Core	\\
NGC3900	&	-1	&	29.4	&	-20.17	&	-19.44	&	7.50	&	37.83	&	37.62	&	$<$36.89	&	L2:	&	S\'{e}rsic	\\
NGC3945	&	-1	&	22.5	&	-20.41	&	-19.68	&	8.05	&	38.53	&	38.19	&	36.8	&	L2	&	Double	\\
NGC3982	&	3	&	17	&	-19.47	&	-17.93	&	6.37	&	40.52	&	39.33	&	$<$36.24	&	S1.9	&	S\'{e}rsic	\\
NGC3998	&	-2	&	14.1	&	-19.26	&	-18.65	&	8.86	&	39.85	&	39.96	&	37.83	&	L1.9	&	Double	\\
NGC4013	&	3	&	17	&	-19.92	&	-18.38	&	6.67	&	37.18	&	37.33	&	$<$36.24	&	T2	&	S\'{e}rsic?	\\
NGC4111	&	-1	&	15	&	-19.28	&	-18.55	&	7.60	&	39.15	&	39.25	&	$<$36.24	&	L2	&	Double?	\\
NGC4125	&	-5	&	23.9	&	-21.22	&	-21.22	&	8.35	&	38.74	&	38.59	&	$<$36.53	&	T2	&	Double	\\
NGC4143	&	-2	&	15.9	&	-19.11	&	-18.5	&	8.17	&	38.47	&	38.73	&	36.7	&	L1.9	&	Double	\\
NGC4150	&	-2	&	13.7	&	-18.29	&	-17.68	&	6.68	&	37.94	&	38	&	$<$36.05	&	T2	&	S\'{e}rsic	\\
NGC4151	&	2	&	14.2	&	-20.05	&	-18.82	&	6.87	&	41.2	&	40.78	&	37.33	&	S1.5	&	S\'{e}rsic	\\
NGC4168	&	-5	&	16.8	&	-19.07	&	-19.07	&	7.98	&	37.69	&	37.91	&	36.72	&	S1.9:	&	Core	\\
NGC4192	&	2	&	16.8	&	-21.11	&	-19.88	&	7.40	&	40.18	&	39.99	&	$<$36.34	&	T2	&	S\'{e}rsic	\\
NGC4203	&	-3	&	15.1	&	-19.29	&	-18.82	&	7.82	&	38.44	&	38.69	&	37.12	&	L1.9	&	S\'{e}rsic	\\
NGC4220	&	-1	&	17	&	-18.93	&	-18.2	&	7.01	&	39.28	&	38.9	&	$<$36.38	&	T2	&	Double?	\\
NGC4261	&	-5	&	31.6	&	-21.14	&	-21.14	&	8.89	&	39.6	&	39.23	&	39.31	&	L2	&	Core	\\
NGC4278	&	-5	&	16.1	&	-20.06	&	-20.06	&	8.59	&	39.26	&	39.35	&	38.14	&	L1.9	&	Core	\\
NGC4281	&	-1	&	35.1	&	-20.57	&	-19.84	&	8.52	&	$<$38.29	&	38.12	&	$<$36.87	&	T2::	&	S\'{e}rsic?	\\
NGC4293	&	0	&	17	&	-20.23	&	-19.37	&	7.12	&	39.4	&	39.21	&	36.38	&	L2	&	Double?	\\
\label{basicprop}
\end{tabular}
\begin{flushleft}
From left to right, columns show: Galaxy name; numerical Hubble type index; distance (Mpc); total B-band absolute magnitude of galaxy; bulge B-band absolute magnitude of galaxy; log black hole mass; log [O{\rm {\sc iii}}] luminosity (erg~s$^{-1}$); log H$\beta$ luminosity (erg~s$^{-1}$); log 15~GHz radio luminosity (erg~s$^{-1}$); AGN optical spectral classification (S:Seyfert, L:LINER, T:transition, numbers denote type 1/2/intermediate, multiple classifications denote intermediate types, : and :: denote uncertain and highly uncertain classifications.  See \citet{hb} for details); nuclear surface brightness profile classification, based on S\'{e}rsic, Core-S\'{e}rsic and Double-S\'{e}rsic fits (see text).  For brightness profile classifications with a question mark, the fits were uncertain.
\end{flushleft}
\end{minipage}
\end{table*}

\begin{table*}
\begin{minipage}{168mm}
\contcaption{Basic properties of the galaxies in our sample for which {\it HST} images were obtained.}
\begin{tabular}{lccccccccll}
\hline
Galaxy & T & D & $M_{B, total}$ & $M_{B, bulge}$ & $\log\left(\frac{M_{BH}}{M_{\sun}}\right)$ & $\log L_{[OIII]}$ & $\log L_{H\beta}$ & $\log L_{\nu, radio}$ & AGN & Class \\
\hline
NGC4314	&	1	&	9.7	&	-18.76	&	-17.74	&	7.19	&	37.94	&	38.09	&	$<$35.75	&	L2	&	S\'{e}rsic	\\
NGC4350	&	-2	&	16.8	&	-19.2	&	-18.59	&	7.95	&	$<$37.99	&	37.78	&	$<$36.18	&	T2::	&	Core	\\
NGC4374	&	-5	&	18.4	&	-21.31	&	-21.31	&	8.88	&	39.15	&	38.93	&	38.57	&	L2	&	Core	\\
NGC4378	&	1	&	35.1	&	-20.51	&	-19.49	&	8.06	&	39.06	&	38.46	&	$<$37.04	&	S2	&	S\'{e}rsic	\\
NGC4388	&	3	&	16.8	&	-20.34	&	-18.8	&	6.77	&	40.78	&	39.79	&	36.8	&	S1.9	&	S\'{e}rsic?	\\
NGC4394	&	3	&	16.8	&	-19.62	&	-18.08	&	7.17	&	38.63	&	38.2	&	$<$36.18	&	L2	&	Double?	\\
NGC4429	&	-1	&	16.8	&	-20.16	&	-19.43	&	8.06	&	38.69	&	38.7	&	$<$36.27	&	T2	&	S\'{e}rsic?	\\
NGC4435	&	-2	&	16.8	&	-19.52	&	-18.91	&	7.70	&	39.02	&	39.2	&	$<$36.27	&	T2/H:	&	Double	\\
NGC4438	&	0	&	16.8	&	-20.64	&	-19.78	&	7.45	&	39.69	&	39.64	&	$<$36.18	&	L1.9	&	S\'{e}rsic?	\\
NGC4450	&	2	&	16.8	&	-20.38	&	-19.15	&	7.44	&	38.34	&	38.3	&	36.66	&	L1.9	&	S\'{e}rsic	\\
NGC4459	&	-1	&	16.1	&	-19.83	&	-19.1	&	7.85	&	37.89	&	38.28	&	$<$36.19	&	T2:	&	S\'{e}rsic	\\
NGC4472	&	-5	&	16.3	&	-21.73	&	-21.73	&	8.79	&	$<$37.78	&	37.07	&	36.81	&	S2::	&	Core	\\
NGC4477	&	-2	&	16.8	&	-19.83	&	-19.22	&	7.91	&	39.12	&	38.57	&	$<$36.23	&	S2	&	S\'{e}rsic?	\\
NGC4486	&	-4	&	16.1	&	-21.54	&	-21.54	&	9.02	&	39.53	&	39.27	&	39.64	&	L2	&	Core	\\
NGC4494	&	-5	&	17.1	&	-20.61	&	-20.61	&	7.57	&	37.6	&	$<$37.54	&	$<$36.15	&	L2::	&	S\'{e}rsic?	\\
NGC4501	&	3	&	16.8	&	-21.27	&	-19.73	&	7.81	&	39.31	&	38.6	&	$<$36.27	&	S2	&	Double?	\\
NGC4550	&	-1.5	&	15.8	&	-18.64	&	-17.97	&	6.75	&	38.22	&	37.96	&	36.02	&	L2	&	Double	\\
NGC4552	&	-5	&	15.3	&	-20.36	&	-20.36	&	8.54	&	38.07	&	38.02	&	37.92	&	T2:	&	Core	\\
NGC4565	&	3	&	17.5	&	-22.11	&	-20.57	&	7.46	&	39.39	&	38.74	&	36.83	&	S1.9	&	S\'{e}rsic?	\\
NGC4569	&	2	&	16.8	&	-21.34	&	-20.11	&	7.46	&	39.83	&	39.79	&	$<$36.27	&	T2	&	Double?	\\
NGC4579	&	3	&	16.8	&	-20.84	&	-19.3	&	7.79	&	39.47	&	39.27	&	37.68	&	S1.9/L1.9	&	S\'{e}rsic?	\\
NGC4589	&	-5	&	22	&	-20.03	&	-20.03	&	8.33	&	39.1	&	38.74	&	37.54	&	L2	&	Core	\\
NGC4596	&	-1	&	16.8	&	-19.63	&	-18.9	&	7.61	&	$<$37.46	&	37.46	&	$<$36.27	&	L2::	&	Double?	\\
NGC4636	&	-5	&	14.7	&	-20.4	&	-20.4	&	8.15	&	37.94	&	38.02	&	36.39	&	L1.9	&	S\'{e}rsic	\\
NGC4698	&	2	&	16.8	&	-19.89	&	-18.66	&	7.61	&	38.88	&	38.25	&	$<$36.23	&	S2	&	S\'{e}rsic	\\
NGC4725	&	2	&	12.4	&	-20.68	&	-19.45	&	7.51	&	38.55	&	37.73	&	$<$35.92	&	S2:	&	Double	\\
NGC4736	&	2	&	5.2	&	-19.83	&	-18.6	&	7.12	&	37.65	&	37.48	&	35.44	&	L2	&	Double	\\
NGC4750	&	2	&	26.1	&	-20.14	&	-18.91	&	7.46	&	38.86	&	38.92	&	$<$36.79	&	L1.9	&	Double?	\\
NGC4772	&	1	&	16.3	&	-19.17	&	-18.15	&	7.61	&	38.49	&	38.46	&	36.73	&	L1.9	&	S\'{e}rsic?	\\
NGC4826	&	2	&	7.5	&	-20.55	&	-19.32	&	6.85	&	39.03	&	38.9	&	$<$35.48	&	T2	&	S\'{e}rsic?	\\
NGC4866	&	-1	&	16	&	-19.29	&	-18.56	&	8.21	&	38.41	&	38.15	&	$<$36.23	&	L2	&	S\'{e}rsic?	\\
NGC5273	&	-2	&	16.5	&	-18.71	&	-18.1	&	6.32	&	39.03	&	39.14	&	$<$36.26	&	S1.5	&	S\'{e}rsic	\\
NGC5322	&	-5	&	31.2	&	-21.43	&	-21.43	&	8.39	&	38.53	&	38.5	&	37.87	&	L2::	&	Core	\\
NGC5377	&	1	&	31	&	-20.52	&	-19.5	&	7.84	&	39.38	&	39.1	&	37.25	&	L2	&	Double	\\
NGC5448	&	1	&	32.6	&	-20.85	&	-19.83	&	7.30	&	39.83	&	39.91	&	$<$36.98	&	L2	&	S\'{e}rsic?	\\
NGC5485	&	-2	&	25.9	&	-19.76	&	-19.15	&	8.19	&	38.06	&	37.65	&	$<$36.78	&	L2:	&	Core	\\
NGC5566	&	2	&	26.4	&	-21.33	&	-20.1	&	7.73	&	38.6	&	38.35	&	$<$36.66	&	L2	&	S\'{e}rsic?	\\
NGC5678	&	3	&	35.6	&	-21.09	&	-19.55	&	7.42	&	39.02	&	39.25	&	$<$36.88	&	T2	&	Double	\\
NGC5746	&	3	&	29.4	&	-22.19	&	-20.65	&	8.13	&	39.06	&	38.76	&	$<$36.71	&	T2	&	Double	\\
NGC5813	&	-5	&	32.2	&	-21.12	&	-21.12	&	8.44	&	38.38	&	38.25	&	37.17	&	L2:	&	Core	\\
NGC5838	&	-3	&	28.5	&	-20.56	&	-20.09	&	8.63	&	39.65	&	39.64	&	36.89	&	T2::	&	S\'{e}rsic	\\
NGC5846	&	-5	&	24.9	&	-21.07	&	-21.07	&	8.42	&	38.52	&	38.44	&	37.37	&	T2:	&	Core?	\\
NGC5866	&	-1	&	15.3	&	-20.1	&	-19.37	&	7.84	&	38.61	&	38.33	&	37.02	&	T2	&	S\'{e}rsic?	\\
NGC5982	&	-5	&		&		&		&	8.44	&		&		&		&		&	S\'{e}rsic	\\
NGC5985	&	3	&		&		&		&	7.71	&		&		&		&		&	S\'{e}rsic?	\\
NGC6340	&	0	&	22	&	-20.04	&	-19.18	&	7.56	&	38.24	&	38.06	&	$<$36.64	&	L2	&	S\'{e}rsic?	\\
NGC6703	&	-2.5	&	26.7	&	-20.16	&	-19.61	&	7.95	&	37.97	&	37.83	&	$<$36.81	&	L2::	&	Double	\\
NGC7177	&	3	&	18.2	&	-19.83	&	-18.29	&	7.30	&	38.93	&	38.95	&	$<$36.34	&	T2	&	S\'{e}rsic?	\\
NGC7217	&	2	&	16	&	-20.49	&	-19.26	&	7.52	&	39.58	&	39.29	&	$<$35.96	&	L2	&	Double	\\
NGC7331	&	3	&	14.9	&	-21.48	&	-19.94	&	7.47	&	38.79	&	38.36	&	$<$36.16	&	T2	&	S\'{e}rsic?	\\
NGC7626	&	-5	&	45.6	&	-21.23	&	-21.23	&	8.69	&	38.57	&	38.36	&	38.7	&	L2::	&	Core	\\
NGC7742	&	3	&	22.2	&	-19.46	&	-17.92	&	6.38	&	39.01	&	38.61	&	$<$36.51	&	T2/L2	&	S\'{e}rsic	\\
NGC7743	&	-1	&	20.7	&	-19.42	&	-18.69	&	6.72	&	40.37	&	39.65	&	36.59	&	S2	&	S\'{e}rsic	\\
NGC7814	&	2	&	13.2	&	-19.57	&	-18.34	&	7.87	&	$<$36.24	&	36.25	&	$<$36.06	&	L2::	&	S\'{e}rsic?	\\
\end{tabular}
\end{minipage}
\end{table*}

The images from WFPC2 and ACS were calibrated using the standard On The Fly Reprocessing (OTFR) system\footnote{http://www.stsci.edu/instruments/wfpc2/Wfpc2\_\\dhb/WFPC2\_longdhbcover.html}$^{,}$\footnote{http://www.stsci.edu/hst/acs/documents/handbooks/\\DataHandbookv5/ACS\_longdhbcover.html}, however the NICMOS images calibrated in this way contained a ``pedestal'', which is a time-variable bias that is different in each quadrant. It is believed that this is caused by small temperature changes, which are different in each quadrant because they use separate amplifiers. To correct for the pedestal the standard CalnicA software that is used in the NICMOS OTFR system~\footnote{http://www.stsci.edu/hst/nicmos/documents/handbooks/\\DataHandbookv7/} was applied using IRAF to the raw images, omitting all of the steps after and including flat field correction. The task BIASEQ in IRAF was then used to equalise the different biases in the 4 quadrants. Finally, CalnicA was again applied using all of the steps after and including flat field correction, and omitting the steps that had already been performed. The final images still contain a bias, but this is now constant over the whole image and so can be accounted for by including a constant background level when fitting the profile model to the image. These NICMOS images also contain several bad pixels --- some of these are cold pixels, which appear dark because they have an abnormally low sensitivity; and some of these are hot pixels, which appear bright because they have an abnormally high dark current. There are also regions, called grots, that appear dark because flecks of antireflective paint scraped off of the optical baffles have reduced their sensitivity. These pixels are identified in the data quality files and are masked in the following analysis. The NIC2 camera also has a coronographic hole for observing targets that are close to bright objects, but when it is not used it still appears in the image as a bright patch, caused by emission from warmer structures that are behind it. This hole moves over time, which results in 2 patches in the image due to using reference files that were taken at a different time. These regions have been masked.

\section{Profile Fitting}

\subsection{Galaxy Models}

We fitted the brightness profiles of all galaxies in our sample with two models. The \citet{sersic} model is parameterised as follows:
\begin{equation}
I(r) = I_{e}\exp\left(-b_{n}\left[\left(\frac{r}{r_{e}}\right)^{1/n} - 1\right]\right)
\end{equation}
This model is described by three parameters: the effective radius $r_{e}$, the intensity at the effective radius $I_{e}$, and the S\'{e}rsic index $n$, which parameterises the curvature of the profile. For $1 \leq n \leq 10$, the quantity $b_{n}$ can be approximated by $b_{n} \approx 1.9992n - 0.3271$ \citep{caon}. We also use the Core-S\'{e}rsic model introduced by \citet{graham03}, which combines a S\'{e}rsic profile at large radii with a power law near the centre:
\begin{equation}
I(r) = I'\left[I + \left(\frac{r_{b}}{r}\right)^{\alpha}\right]^{\gamma/\alpha}\exp\left[-b_{n}\left(\frac{r^{\alpha} + r_{b}^{\alpha}}{r_{e}^{\alpha}}\right)^{1/\alpha n}\right]
\end{equation}
This model uses six parameters, including the S\'{e}rsic index $n$ and effective radius $r_{e}$ of the outer S\'{e}rsic profile. The inner power law profile has a logarithmic slope $\gamma$, and the break radius $r_{b}$ is the radius separating the inner power law and outer S\'{e}rsic profiles. The parameter $\alpha$ determines the sharpness of the transition between the two profiles, with a high value of $\alpha$ indicating a sharp transition. The intensity $I'$ is related to the intensity at the break radius, $I_{b}$, as follows:
\begin{equation}
I' = I_{b} 2^{-\gamma/\alpha}\exp\left[b_{n}\left(2^{1/\alpha}\frac{r_{b}}{r_{e}}\right)^{1/n}\right]
\end{equation}
The quantity $b_{n}$ is calculated as in the S\'{e}rsic model.

As noted in section 2.1 we excluded late type spiral galaxies from our sample as their brightness profiles are dominated by the disc, however there are still several early type spiral galaxies in our sample. It is possible that there is still a significant contribution from the disc component in the central surface brightness profiles of these galaxies, so for all of the spiral and lenticular galaxies in our sample we also added an exponential disc component to the model.

The profiles of several galaxies in our sample showed an unresolved nuclear source due to the AGN. For these galaxies a central point source was added to the S\'{e}rsic and Core-S\'{e}rsic models.

\citet{cote07} demonstrate that low and intermediate luminosity early-type galaxies often contain a resolved nuclear source, for example a compact spheroidal or flattened stellar component. They model the brightness profiles of these galaxies using a Double-S\'{e}rsic model:
\begin{align}
\nonumber I(r) = &I_{e, 1} \exp\left(-b_{n1}\left[\left(\frac{r}{r_{e, 1}}\right)^{1/n1} - 1\right]\right) \\ 
&+ I_{e, 2} \exp\left(-b_{n2}\left[\left(\frac{r}{r_{e, 2}}\right)^{1/n2} - 1\right]\right)
\end{align}
The inner S\'{e}rsic model (parameterised by $I_{e,1}$, $r_{e, 1}$ and $n1$) describes the resolved nuclear component while the outer S\'{e}rsic profile (parameterised by $I_{e,2}$, $r_{e, 2}$ and $n2$) describes the host galaxy. The brightness profiles of several galaxies in our sample showed a central excess above a single S\'{e}rsic profile that was too extended to be described by a point source, so these galaxies were also fitted with this Double S\'{e}rsic model.

Using the fits with the S\'{e}rsic and Core-S\'{e}rsic laws, we classified a galaxy as `Core' if it satisfied the following criteria \citep[based on those used by][]{trujillo}:
\begin{enumerate}
\item The Core-S\'{e}rsic fit must give a lower reduced $\chi^{2}$ than the S\'{e}rsic fit. We note that \citet{trujillo} required that the reduced $\chi^{2}$ of the Core-S\'{e}rsic fit must be at least a factor of 2 less than that of the S\'{e}rsic fit, however we found that very few galaxies in our sample satisfied such a strict criterion. This discrepancy is likely due to differences in how the data points are weighted in the fits, as they use equal weights whereas we weight the data points using their Poissonian errors, which are larger at smaller radii.
\item To ensure that the core is well sampled by the data the break radius $r_{b}$ must be greater than the radius of the second data point.
\item The slope of the inner power law profile, $\gamma$, must be less than the logarithmic slope of the best-fit S\'{e}rsic model within the break radius.
\end{enumerate}

Galaxies for which the Double-S\'{e}rsic model gave the lowest reduced $\chi^{2}$ were classified as `Double' galaxies (i.e. contain a resolved nuclear source). All other galaxies were classified as `S\'{e}rsic' galaxies.

\subsection{Extracting 1D Profiles}

The 1D semi-major axis brightness profiles of each galaxy were extracted using the ELLIPSE task in IRAF, which fits elliptical isophotes to the galaxy images. Regions of dust that were visible in the images were masked from these fits. The position angle and ellipticity were free to vary with each isophote, enabling radial variations in these quantities, such as isophotal twists, to be accounted for. 

\subsection{Fitting Procedure}

The S\'{e}rsic, Core-S\'{e}rsic and (where necessary) Double-S\'{e}rsic models were fitted to the 1D brightness profiles by minimising the $\chi^{2}$ statistic using the {\it Sherpa}\footnote{http://cxc.harvard.edu/sherpa/} modelling and fitting software package, which is a part of the {\it Chandra} Interactive Analysis of Observations (CIAO\footnote{http://cxc.harvard.edu/ciao/}) software. The data points were weighted in these fits using their Poissonian errors.

To accurately reproduce the brightness profile at small radii it is necessary to account for the point spread function (psf) of the instrument used. For each image we created a 2D model of the psf at the position of the galaxy's centre on the detector using the Tiny Tim software\footnote{http://www.stsci.edu/software/tinytim/tinytim.html}. It was necessary to create a separate psf model for each image because the psf depends on the position on the detector. The ELLIPSE task in IRAF was used to extract the 1D profile of the psf, then the galaxy model was convolved with this psf in {\it Sherpa} before fitting to the observed brightness profile.

In the Core-S\'{e}rsic model the parameters $\alpha$, n and $\gamma$ tend to be degenerate. \citet{trujillo} recommend that $\alpha$ should be fixed at a large value to force an abrupt transition between the outer S\'{e}rsic and inner power-law profiles, however we found that in several galaxies the profile was poorly fitted by such a fixed-$\alpha$ model, but it was well fitted if $\alpha$ was free to vary. Therefore we fixed $\alpha$ at $\alpha=50$ to obtain an initial fit and then we allowed $\alpha$ to vary freely to obtain the final fit. We investigated how robust the parameters were for the free-$\alpha$ fits and we found that the parameters of the outer S\'{e}rsic component were uncertain in some of the galaxies, particularly those with a low value of $\alpha$. This may also be caused by the limited radial extent of the fitting regions, which were comparable to the effective radius $r_{e}$ in some galaxies. However, the uncertainties in $\gamma$ were relatively small, and the break radii $r_{b}$ of the Core-S\'{e}rsic models were well within the fitting region, so the classifications as Core or S\'{e}rsic were robust.

\subsection{Results}

\begin{figure*}
\mbox{
	\includegraphics[width=56mm]{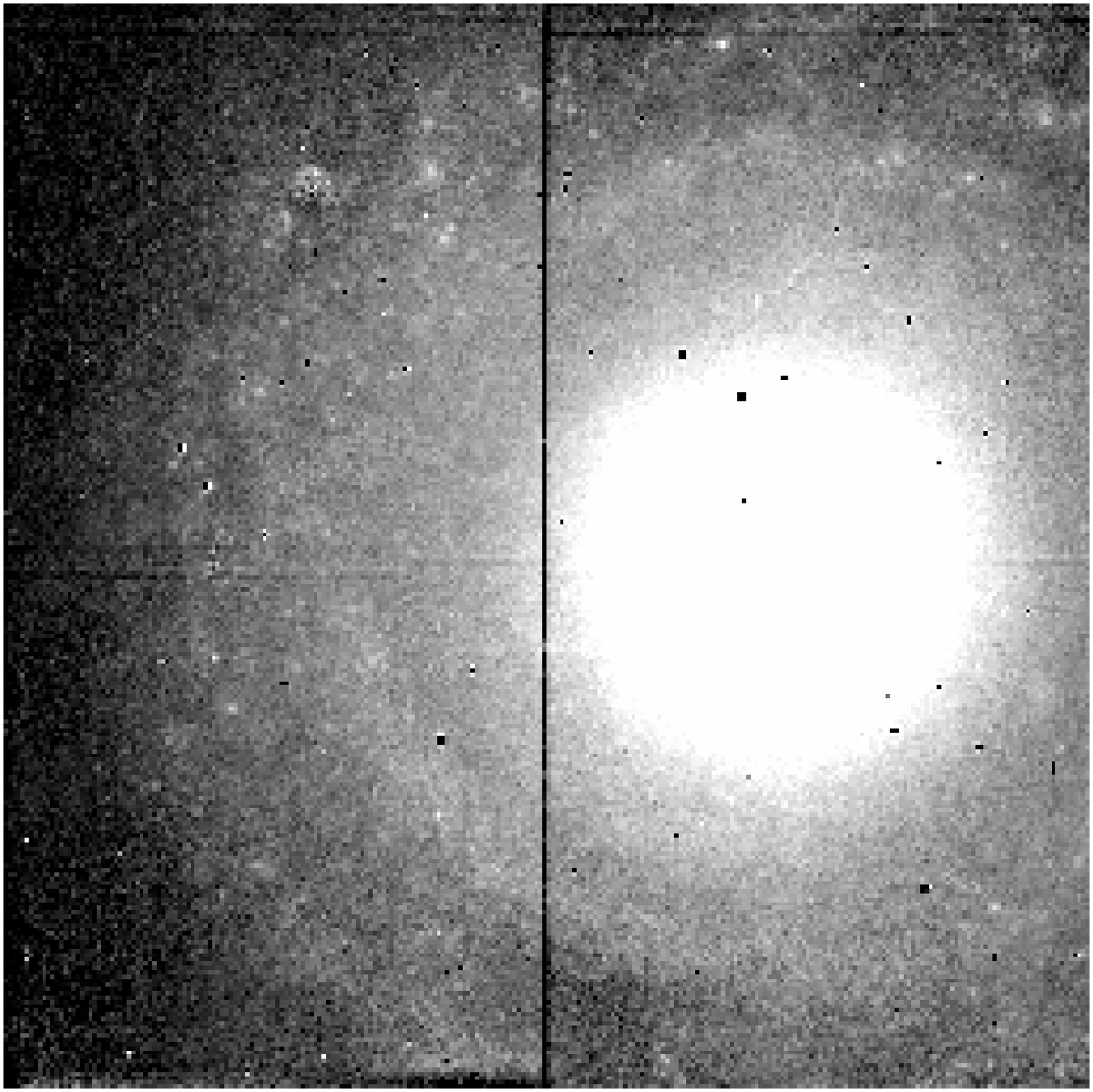}
	\includegraphics[width=56mm]{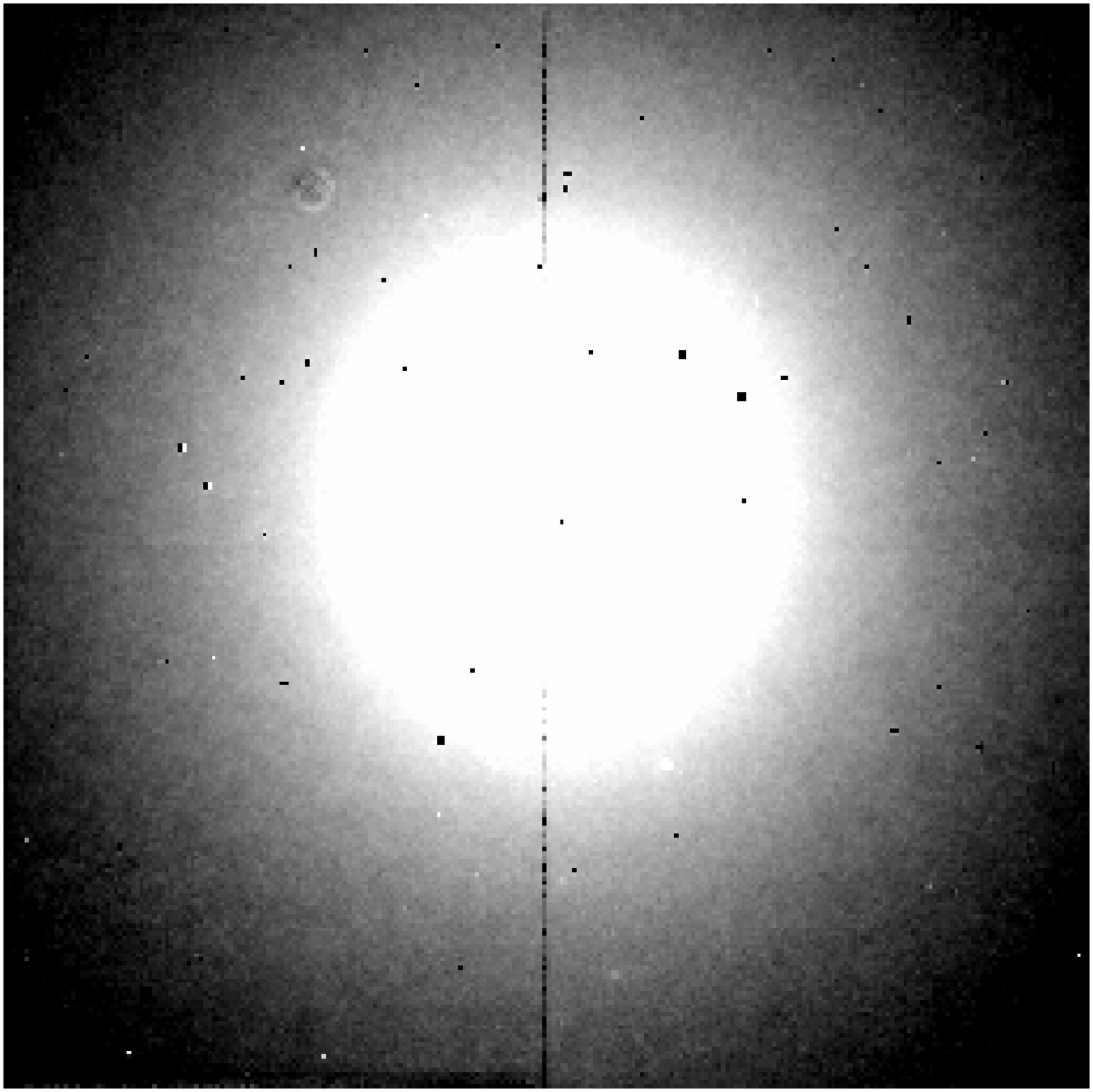}
	\includegraphics[width=56mm]{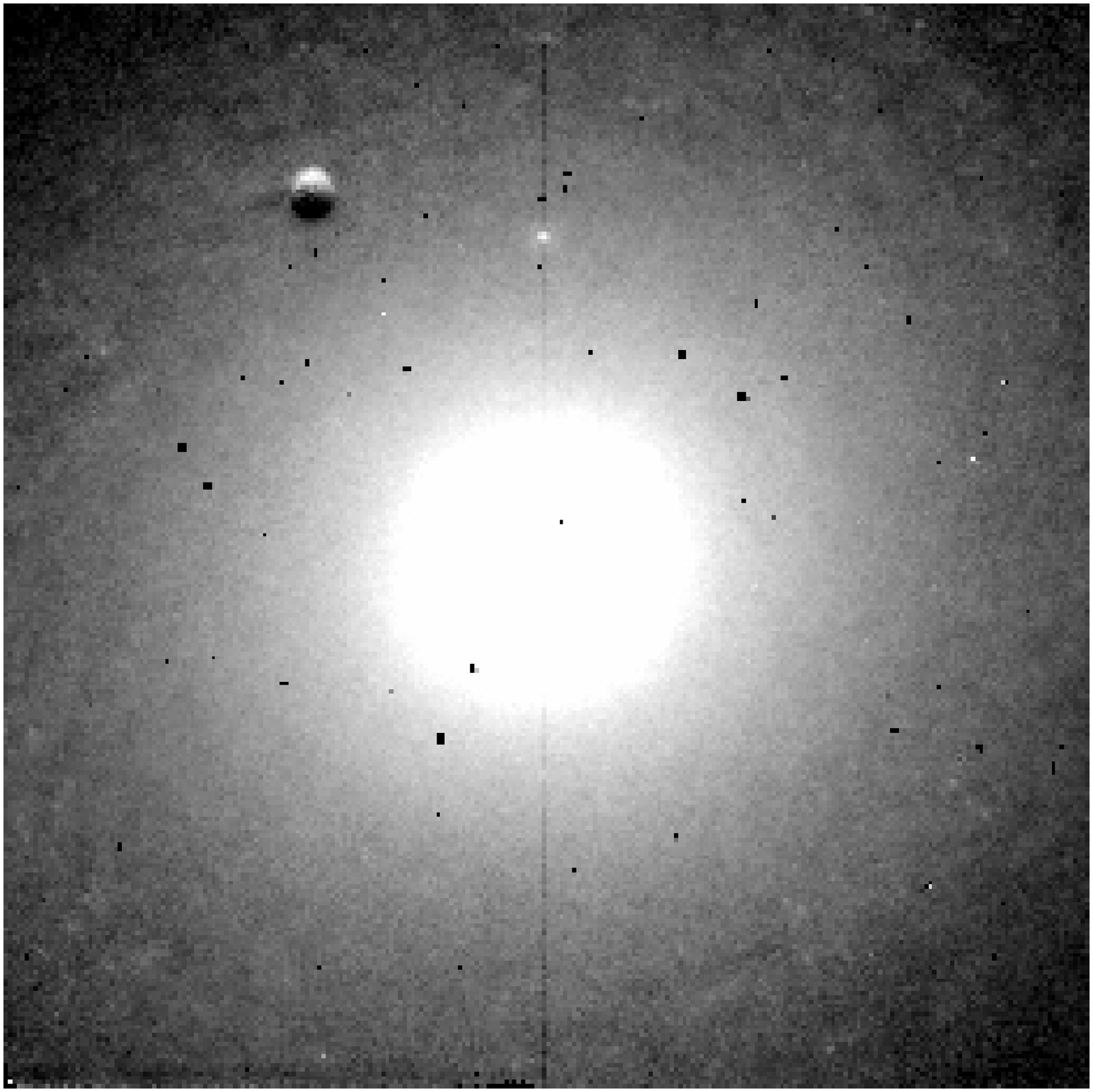}}
\mbox{
	\includegraphics[width=56mm]{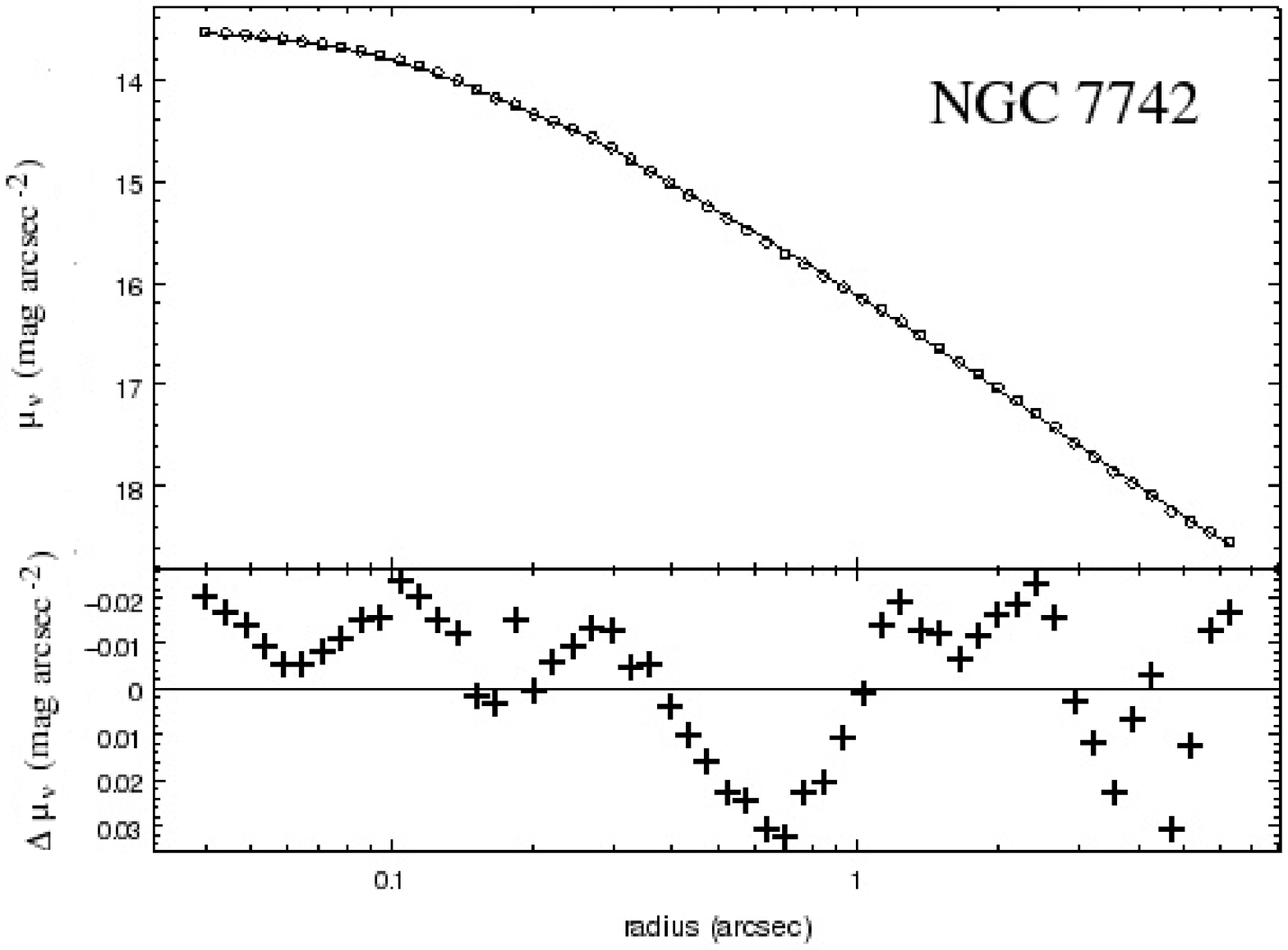}
	\includegraphics[width=56mm]{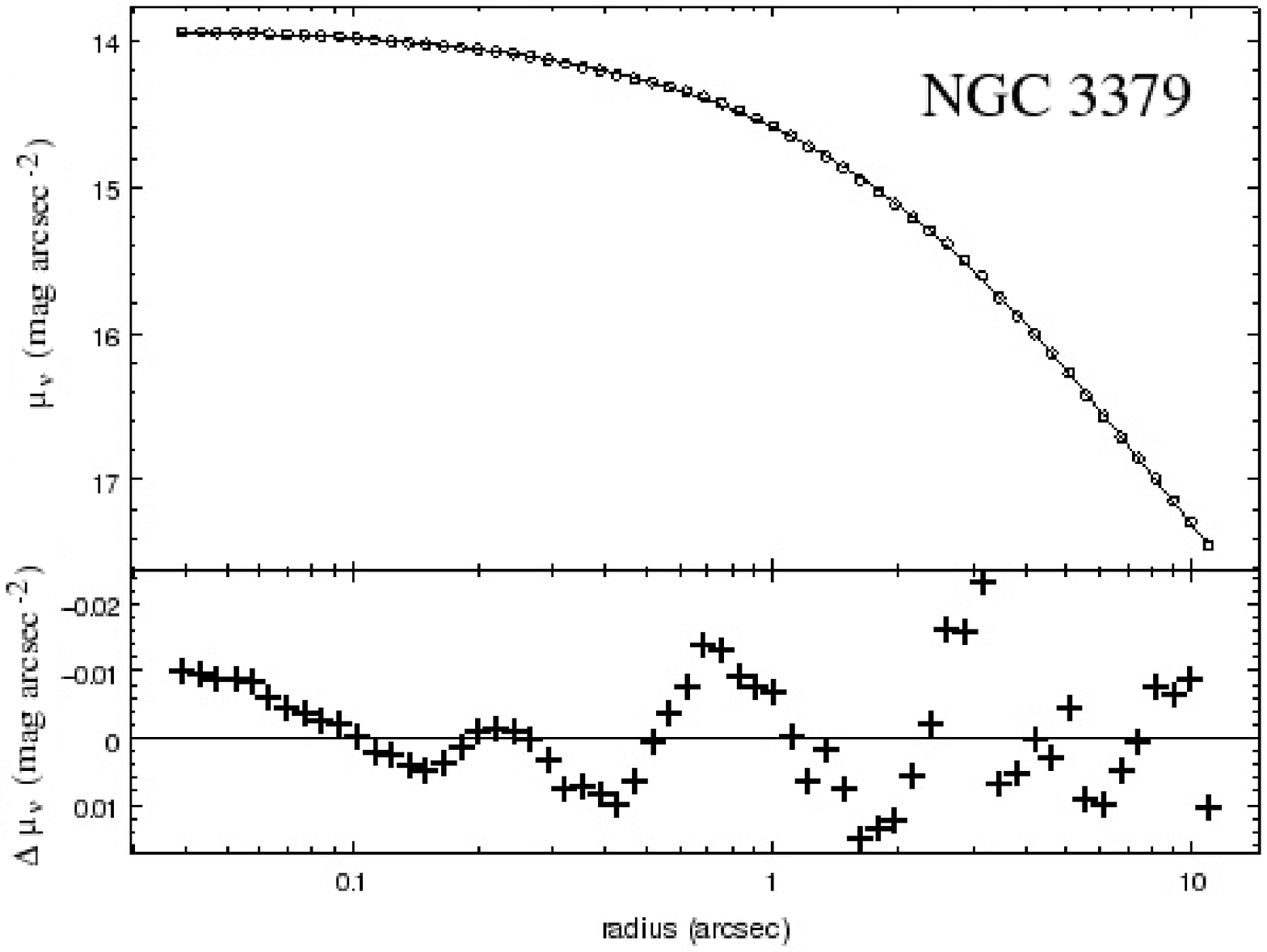}
	\includegraphics[width=56mm]{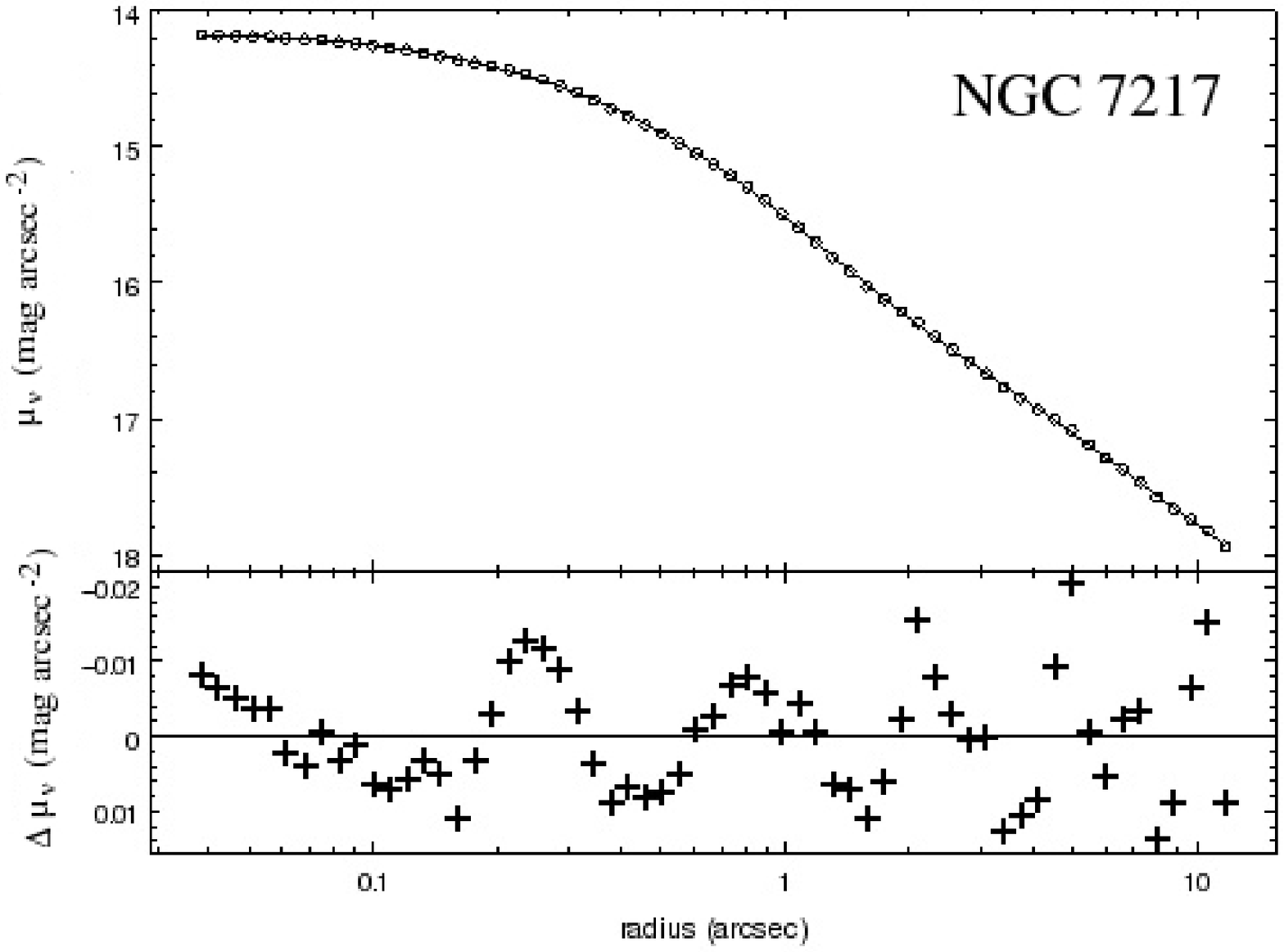}}
\caption{Galaxy images (top row) and radial brightness profiles (bottom row) for a confident S\'{e}rsic fit (NGC7742; left), Core fit (NGC 3379; centre) and Double S\'{e}rsic fit (NGC7217; right).}
\label{confident_fits}
\end{figure*}

\begin{figure*}
\mbox{
	\includegraphics[width=56mm]{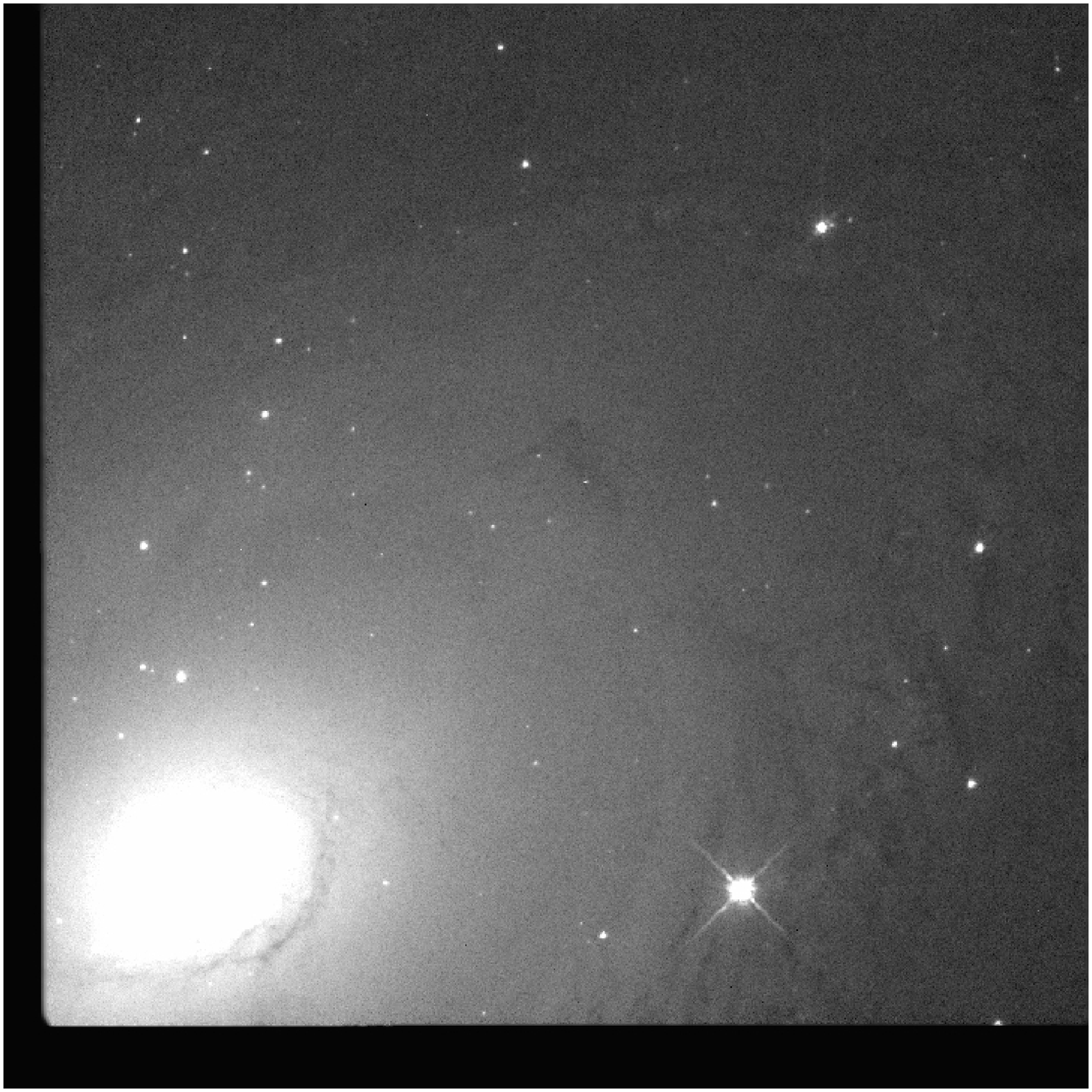}
	\includegraphics[width=56mm]{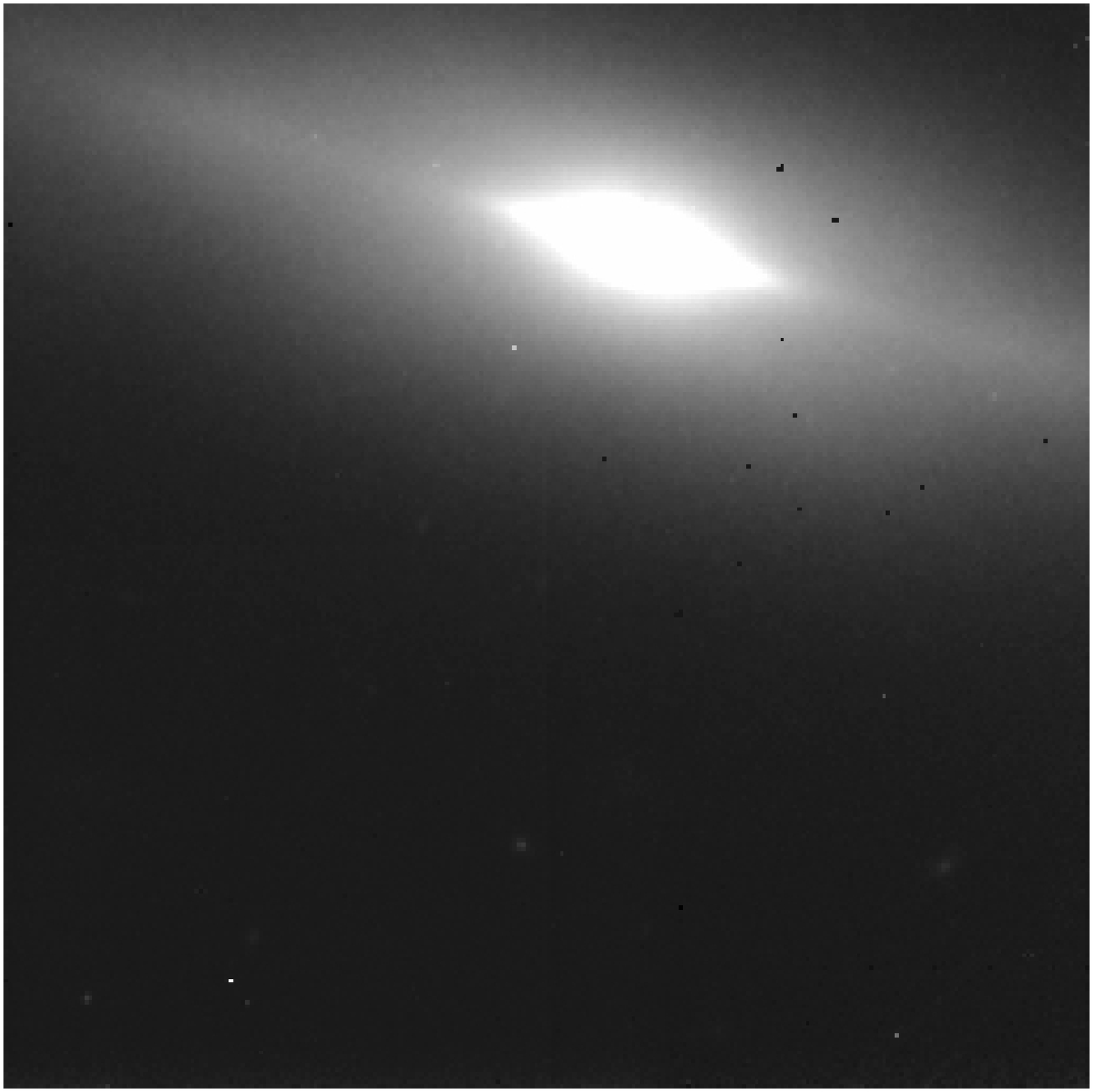}}
\mbox{
	\includegraphics[width=56mm]{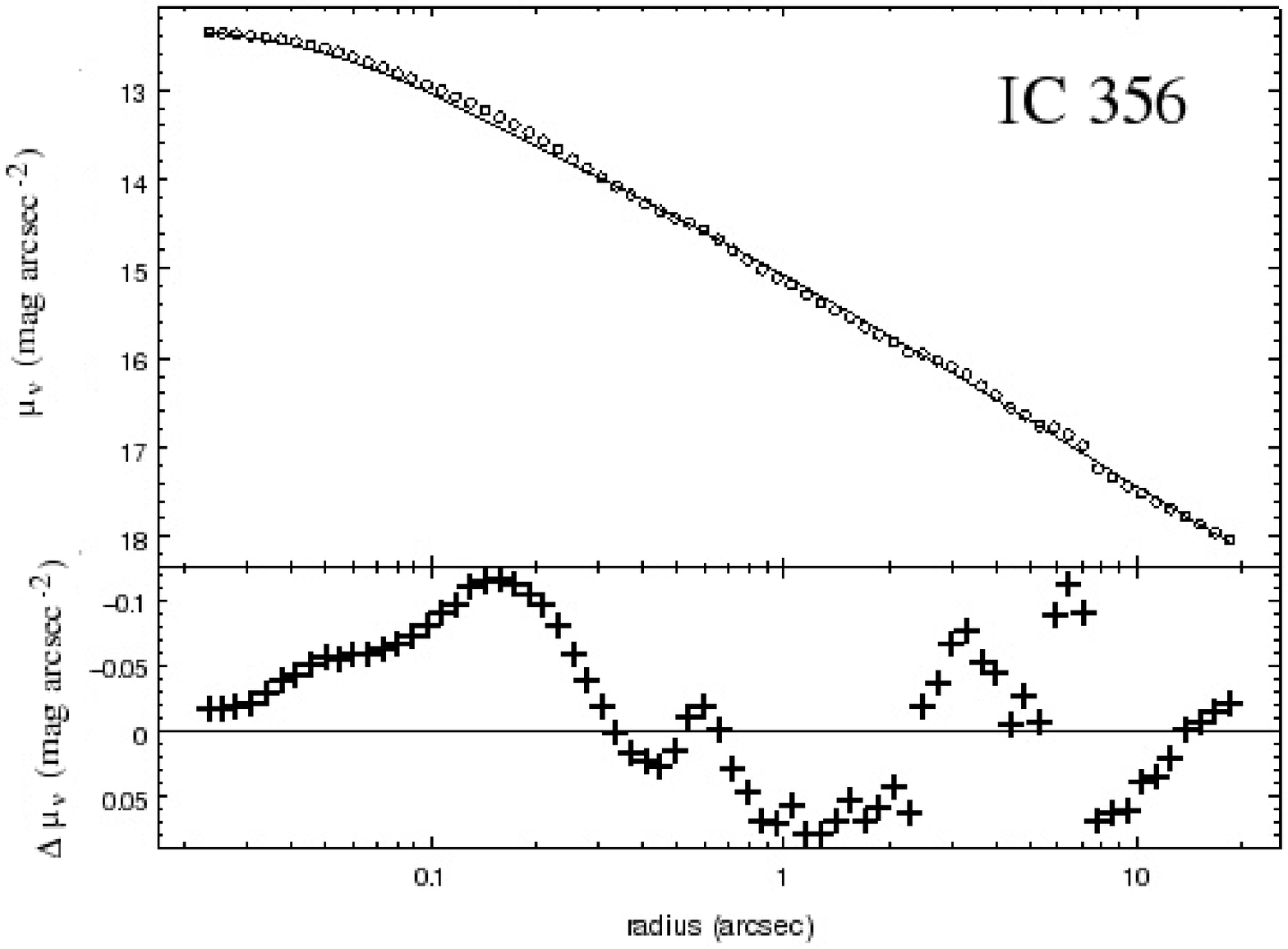}
	\includegraphics[width=56mm]{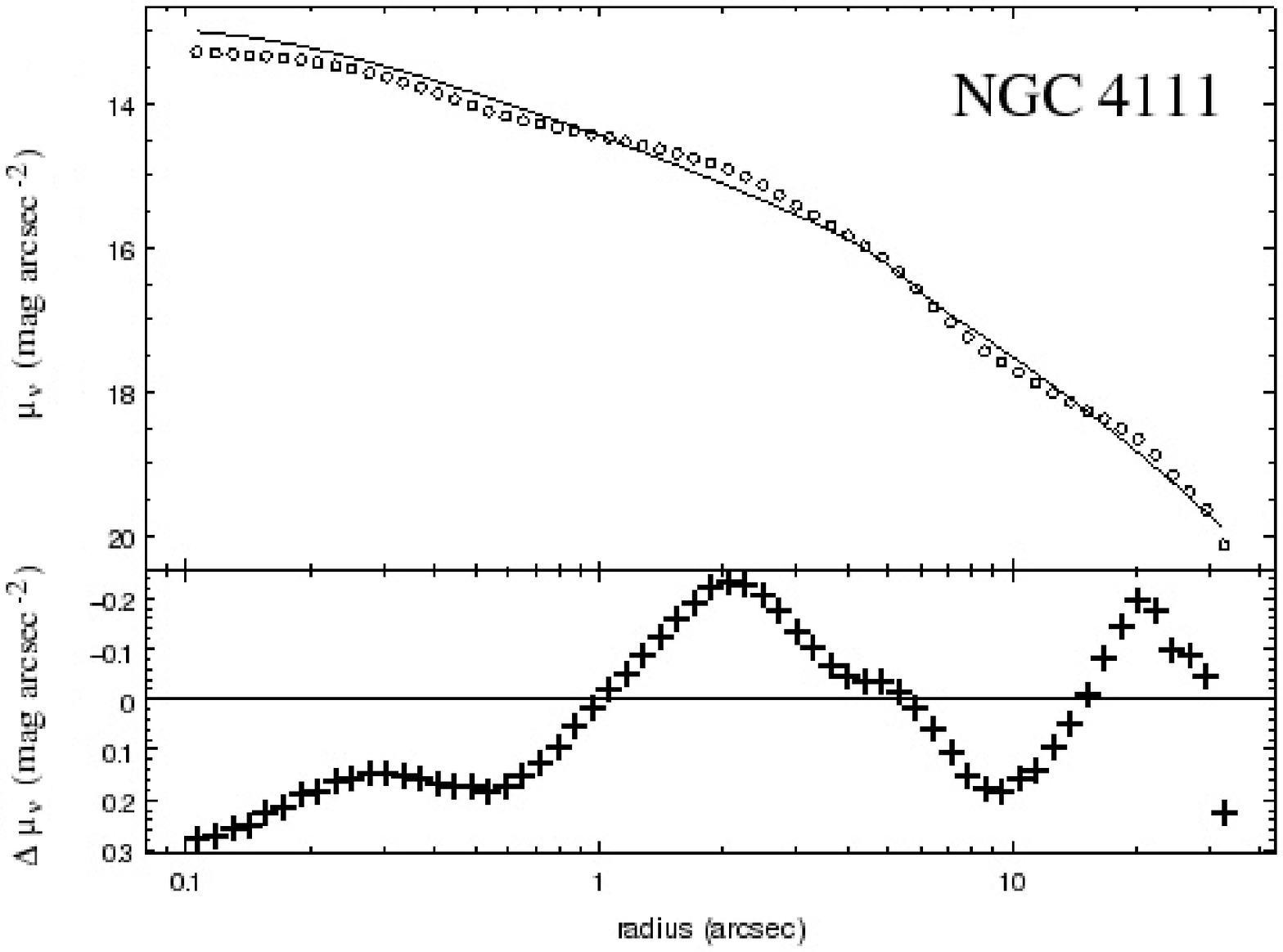}}
\caption{Galaxy images (top row) and radial brightness profiles (bottom row) for a fit that is uncertain due to dust (IC 356; left) and a fit that is uncertain because the galaxy has a complex profile (NGC 4111; right).}
\label{uncertain_fits}
\end{figure*}

We were able to obtain fits for all 110 galaxies in our sample, however we are only confident in 62 of these fits, which are reported in table~\ref{paramconf}. 44 of the remaining galaxies showed large regions of dust that made the fits uncertain, even after we had masked most of this dust. The profiles of the remaining 4 galaxies could not be well described by a S\'{e}rsic, Core-S\'{e}rsic or Double-S\'{e}rsic model. We report these fits in table~\ref{paramuncert}, however they will be discarded for most of the following analysis. We present examples of confident fits using the S\'{e}rsic, Core-S\'{e}rsic and Double-S\'{e}rsic models in fig.~\ref{confident_fits} and two uncertain fits in fig.~\ref{uncertain_fits}, showing for each the galaxy image and the radial profiles of the galaxy and the best fitting model.

\begin{table*}
\begin{minipage}{160mm}
\caption{Best fit parameters of the Core, S\'{e}rsic and Double S\'{e}rsic galaxies from the confident fits.}
\label{paramconf}
\begin{tabular}{llcccccccc}
\hline
\multicolumn{10}{c}{Core Galaxies} \\
Galaxy & Instrument & Filter & $\mu'$\footnote{Corrected for extinction using values from \citet{hb} and converted to the Johnson V band, assuming that the galaxy has the spectrum of a K2 giant.} & $r_{b}$\footnote{Measured in arcsec.} & $\gamma$ & $n$ & $r_{e}^{b}$ & $\alpha$ & $m_{pt}$\footnote{Point source magnitude, corrected for extinction using values from \citet{hb} and converted to the Johnson V band, assuming a spectral index of 1.} \\
\hline
NGC524 	&	NICMOS	&	F160W	&	13.32	&	0.18	&	0.11	&	2.80	&	14.88	&	134.14	&	 - 	\\
NGC2832	&	NICMOS	&	F160W	&	13.51	&	0.64	&	0.11	&	2.90	&	9.03	&	4.69	&	 - 	\\
NGC2841	&	NICMOS	&	F160W	&	10.05	&	0.17	&	0.35	&	5.03	&	77.48	&	102.27	&	 - 	\\
NGC3193	&	WFPC2	&	F702W	&	5.39	&	0.26	&	0.19	&	9.36	&	302.17	&	1.54	&	 - 	\\
NGC3379	&	NICMOS	&	F160W	&	10.89	&	1.19	&	0.06	&	2.90	&	6.93	&	1.3	&	  - 	\\
NGC3607	&	NICMOS	&	F160W	&	13.38	&	0.22	&	0.00	&	1.93	&	7.63	&	1.00	&	 - 	\\
NGC3608	&	WFPC2	&	F814W	&	9.00	&	0.31	&	0.29	&	6.65	&	71.59	&	4.01	&	 - 	\\
NGC3898	&	NICMOS	&	F160W	&	9.44	&	0.22	&	0.27	&	4.79	&	17.77	&	4.60	&	 - 	\\
NGC4168	&	WFPC2 	&	F702W	&	14.63	&	0.49	&	0.00	&	3.67	&	41.84	&	1.30	&	23.05	\\
NGC4261	&	NICMOS	&	F160W	&	11.30	&	1.55	&	0.00	&	3.10	&	7.06	&	1.87	&	21.50	\\
NGC4278	&	NICMOS	&	F160W	&	9.26	&	0.81	&	0.13	&	5.38	&	48.56	&	3.22	&	22.90	\\
NGC4350	&	NICMOS	&	F160W	&	7.19	&	0.55	&	0.25	&	5.77	&	18.99	&	2.22	&	 - 	\\
NGC4374	&	NICMOS	&	F160W	&	11.20	&	1.86	&	0.12	&	3.20	&	12.36	&	1.88	&	20.64	\\
NGC4472	&	NICMOS	&	F160W	&	10.86	&	1.76	&	0.00	&	4.35	&	52.52	&	2.11	&	24.33	\\
NGC4486	&	WFPC2	&	F814W	&	10.89	&	6.68	&	0.27	&	5.32	&	117.19	&	3.38	&	16.90	\\
NGC4552	&	NICMOS	&	F160W	&	11.67	&	0.36	&	0.14	&	2.71	&	6.98	&	7.73	&	 - 	\\
NGC4589	&	NICMOS	&	F160W	&	10.57	&	0.14	&	0.00	&	4.64	&	21.19	&	2.12	&	 - 	\\
NGC5322	&	ACS	&	F814W	&	5.73	&	1.23	&	0.41	&	8.14	&	93.09	&	2.62	&	 - 	\\
NGC5485	&	ACS	&	F814W	&	16.14	&	0.32	&	0.13	&	2.74	&	19.79	&	113.48	&	 - 	\\
NGC5813	&	WFPC2	&	F814W	&	6.06	&	0.74	&	0.16	&	8.86	&	215.30	&	3.88	&	 - 	\\
NGC7626	&	NICMOS	&	F160W	&	9.26	&	0.47	&	0.30	&	5.39	&	19.73	&	2.79	&	 - 	\\
\hline
\multicolumn{10}{c}{S\'{e}rsic Galaxies} \\
Galaxy & Instrument & Filter & $\mu_{e}^{a}$ & $n$ & $r_{e}^{b}$ & $\mu_{e, disc}$ & $r_{e, disc}$ & $m_{pt}^{c}$ \\
\hline
NGC474	&	NICMOS	&	F160W	&	18.12	&	2.09	&	4.22	&	-	&	-	&	20.37	\\
NGC488 	&	NICMOS	&	F160W	&	20.26	&	3.53	&	22.83	&	-	&	-	&	22.05	\\
NGC2685	&	WFPC2	&	F814W	&	20.93	&	3.91	&	29.76	&	-	&	-	&	 - 	\\
NGC2985	&	NICMOS	&	F160W	&	19.50	&	3.45	&	16.52	&	-	&	-	&	20.77	\\
NGC3227	&	NICMOS	&	F160W	&	17.80	&	4.07	&	3.63	&	-	&	-	&	17.89	\\
NGC3245	&	NICMOS	&	F160W	&	17.59	&	3.03	&	6.02	&	-	&	-	&	19.52	\\
NGC3414	&	WFPC2 	&	F814W	&	22.38	&	5.95	&	48.81	&	-	&	-	&	 - 	\\
NGC3516	&	NICMOS	&	F160W	&	16.64	&	1.43	&	2.24	&	-	&	-	&	17.17	\\
NGC3718	&	NICMOS	&	F160W	&	15.95	&	1.50	&	1.04	&	18.05	&	4.78	&	19.21	\\
NGC3900	&	NICMOS	&	F160W	&	22.22	&	6.55	&	38.11	&	-	&	-	&	 - 	\\
NGC3982	&	NICMOS	&	F160W	&	19.60	&	2.47	&	4.84	&	-	&	-	&	20.17	\\
NGC4150	&	NICMOS	&	F160W	&	18.21	&	3.10	&	4.52	&	-	&	-	&	 - 	\\
NGC4151	&	NICMOS	&	F160W	&	18.76	&	4.30	&	9.00	&	-	&	-	&	15.13	\\
NGC4192	&	NICMOS	&	F160W	&	16.76	&	1.93	&	5.27	&	20.76	&	369.78	&	15.51	\\
NGC4203	&	ACS	&	F814W	&	21.49	&	4.30	&	22.92	&	-	&	-	&	21.22	\\
NGC4314	&	NICMOS	&	F160W	&	22.90	&	5.41	&	154.78	&	-	&	-	&	 - 	\\
NGC4378	&	WFPC2	&	F606W	&	19.75	&	2.78	&	5.93	&	21.51	&	24.46	&	 - 	\\
NGC4450	&	WFPC2	&	F814W	&	25.61	&	8.01	&	502.69	&	-	&	-	&	 - 	\\
NGC4459	&	NICMOS	&	F160W	&	18.12	&	3.56	&	8.62	&	-	&	-	&	 - 	\\
NGC4636	&	NICMOS	&	F160W	&	17.87	&	1.20	&	7.21	&	-	&	-	&	 - 	\\
NGC4698	&	WFPC2	&	F814W	&	16.65	&	2.03	&	1.64	&	18.81	&	11.65	&	20.60	\\
NGC5273	&	NICMOS	&	F160W	&	17.85	&	1.66	&	2.08	&	-	&	-	&	19.90	\\
NGC5838	&	NICMOS	&	F160W	&	20.50	&	5.21	&	37.30	&	-	&	-	&	 - 	\\
NGC5982	&	NICMOS	&	F160W	&	17.86	&	2.03	&	5.54	&	-	&	-	&	 - 	\\
NGC7742	&	NICMOS	&	F160W	&	19.94	&	4.64	&	9.55	&	-	&	-	&	 - 	\\
NGC7743	&	NICMOS	&	F160W	&	20.17	&	6.34	&	9.66	&	-	&	-	&	 - 	\\
\end{tabular}
\end{minipage}
\end{table*}

\begin{table*}
\begin{minipage}{160mm}
\contcaption{Best fit parameters of the Core, S\'{e}rsic and Double S\'{e}rsic galaxies from the confident fits.}
\begin{tabular}{llcccccccccc}
\hline
\multicolumn{10}{c}{Double S\'{e}rsic Galaxies} \\
Galaxy & Instrument & Filter & $\mu_{e1}^{a}$ & $n_{1}$ & $r_{e1}^{b}$ & $\mu_{e2}^{a}$ & $n_{2}$ & $r_{e2}^{b}$ & $\mu_{e, disc}$ & $r_{e, disc}$ & $m_{pt}^{c}$ \\
\hline
NGC404	&	NICMOS	&	F160W	&	19.74	&	15.72	&	2.55	&	24.97	&	6.35	&	712.26	&	-	&	-	&	 - 	\\
NGC3169	&	NICMOS	&	F160W	&	14.90	&	0.81	&	0.38	&	17.24	&	1.60	&	6.72	&	-	&	-	&	 - 	\\
NGC3945	&	WFPC2	&	F814W	&	17.12	&	1.57	&	1.10	&	18.67	&	1.04	&	9.33	&	-	&	-	&	20.60	\\
NGC3998	&	ACS	&	F814W	&	17.11	&	3.04	&	0.35	&	19.66	&	2.90	&	9.75	&	-	&	-	&	 - 	\\
NGC4125	&	NICMOS	&	F160W	&	16.55	&	1.04	&	0.84	&	18.46	&	2.32	&	17.56	&	-	&	-	&	 - 	\\
NGC4143	&	NICMOS	&	F160W	&	18.24	&	0.48	&	3.25	&	17.69	&	2.51	&	5.59	&	-	&	-	&	21.12	\\
NGC4435	&	NICMOS	&	F160W	&	16.12	&	0.59	&	0.39	&	16.94	&	1.64	&	5.52	&	-	&	-	&	19.56	\\
NGC4550	&	NICMOS	&	F160W	&	16.96	&	0.16	&	1.92	&	19.12	&	3.56	&	5.10	&	-	&	-	&	 - 	\\
NGC4725	&	NICMOS	&	F160W	&	15.20	&	1.06	&	0.13	&	18.70	&	2.83	&	14.10	&	-	&	-	&	 - 	\\
NGC4736	&	NICMOS	&	F160W	&	16.14	&	0.56	&	8.25	&	17.98	&	3.54	&	19.42	&	-	&	-	&	 - 	\\
NGC5377	&	NICMOS	&	F160W	&	15.06	&	0.95	&	0.26	&	16.91	&	0.63	&	2.81	&	20.41	&	440.82	&	 - 	\\
NGC5678	&	NICMOS	&	F160W	&	15.68	&	0.16	&	0.02	&	23.47	&	6.07	&	133.73	&	22.11	&	367.00	&	 - 	\\
NGC5746	&	NICMOS	&	F160W	&	17.44	&	0.16	&	1.41	&	17.56	&	2.09	&	6.74	&	19.06	&	623.32	&	20.61	\\
NGC6703	&	WFPC2	&	F814W	&	20.57	&	0.82	&	10.87	&	21.95	&	6.22	&	25.78	&	-	&	-	&	 - 	\\
NGC7217	&	NICMOS	&	F160W	&	16.52	&	0.93	&	0.79	&	19.60	&	2.64	&	20.05	&	20.51	&	167.78	&	 - 	\\
\end{tabular}
\end{minipage}
\end{table*}

\begin{table*}
\begin{minipage}{160mm}
\caption{Best fit parameters of the Core, S\'{e}rsic and Double S\'{e}rsic galaxies from the uncertain fits.}
\label{paramuncert}
\begin{tabular}{llccccccc@{\hspace{0.03in}}c@{\hspace{0.03in}}c@{\hspace{0.03in}}c@{\hspace{0.03in}}c}
\hline
\multicolumn{11}{c}{Core Galaxies} \\
Galaxy & Instrument & Filter & $\mu'$\footnote{Corrected for extinction using values from \citet{hb} and converted to the Johnson V band, assuming that the galaxy has the spectrum of a K2 giant.} & $r_{b}$\footnote{Measured in arcsec.} & $\gamma$ & $n$ & $r_{e}^{b}$ & $\alpha$ & $m_{pt}$\footnote{Point source magnitude, corrected for extinction using values from \citet{hb} and converted to the Johnson V band, assuming a spectral index of 1.} & Uncertainty \\
\hline
NGC315	&	WFPC2	&	F814W	&	12.51	&	0.89	&	0.00	&	5.06	&	100.29	&	3.27	&	21.27	&	Dusty	\\
NGC5846	&	WFPC2	&	F702W	&	11.23	&	1.24	&	0.00	&	6.54	&	419.02	&	4.40	&	 - 	&	Dusty	\\
\hline
\multicolumn{11}{c}{S\'{e}rsic Galaxies} \\
Galaxy & Instrument & Filter & $\mu_{e}^{a}$ & $n$ & $r_{e}^{b}$ & $\mu_{e, disc}$ & $r_{e, disc}$ & $m_{pt}^{c}$ & Uncertainty \\
\hline
IC356	&	WFPC2	&	F814W	&	25.83	&	9.16	&	1491.05	&	21.32	&	2904.34	&	 - 	&	Dusty	\\
NGC660	&	ACS	&	F814W	&	19.25	&	0.23	&	8.98	&	-	&	-	&	 - 	&	Dusty	\\
NGC2273	&	NICMOS	&	F160W	&	16.17	&	0.74	&	1.82	&	19.55	&	15.45	&	18.76	&	Dusty	\\
NGC2681	&	NICMOS	&	F160W	&	23.12	&	12.35	&	103.2	&	-	&	-	&	 - 	&	Complex	\\
NGC2683	&	NICMOS	&	F160W	&	22.79	&	6.17	&	184.16	&	19.43	&	369.32	&	 - 	&	Dusty	\\
NGC3190	&	ACS	&	F814W	&	18.85	&	1.97	&	9.54	&	-	&	-	&	 - 	&	Dusty	\\
NGC3368	&	NICMOS	&	F160W	&	17.89	&	2.42	&	14.26	&	-	&	-	&	16.81	&	Dusty	\\
NGC3489	&	WFPC2	&	F814W	&	19.51	&	4.22	&	15.24	&	-	&	-	&	 - 	&	Dusty	\\
NGC3507	&	WFPC2	&	F606W	&	21.01	&	4.8	&	7.95	&	21.75	&	43.06	&	 - 	&	Dusty	\\
NGC3623	&	WFPC2	&	F814W	&	21.16	&	5.81	&	63.13	&	19.68	&	579.52	&	 - 	&	Dusty	\\
NGC3626	&	NICMOS	&	F160W	&	16.67	&	2.32	&	3.14	&	-	&	-	&	17.87	&	Dusty	\\
NGC3627	&	NICMOS	&	F160W	&	17.79	&	2.78	&	10.46	&	19.57	&	344.24	&	 - 	&	Dusty	\\
NGC3628	&	NICMOS	&	F160W	&	25.21	&	5.6	&	5850.16	&	-	&	-	&	 - 	&	Dusty	\\
NGC4013	&	WFPC2	&	F814W	&	19.37	&	0.18	&	7.9	&	20.55	&	228.1	&	 - 	&	Dusty	\\
NGC4281	&	WFPC2	&	F606W	&	25.97	&	8.71	&	509.97	&	-	&	-	&	 - 	&	Dusty	\\
NGC4388	&	NICMOS	&	F160W	&	24.09	&	7.79	&	121.02	&	-	&	-	&	 - 	&	Dusty	\\
NGC4429	&	WFPC2	&	F606W	&	22.78	&	4.52	&	134.39	&	-	&	-	&	17.43	&	Dusty	\\
NGC4438	&	WFPC2	&	F814W	&	18.58	&	3.42	&	12.99	&	-	&	-	&	 - 	&	Dusty	\\
NGC4477	&	WFPC2	&	F606W	&	21.3	&	3.58	&	35.93	&	-	&	-	&	 - 	&	Dusty	\\
NGC4494	&	WFPC2	&	F814W	&	33.22	&	17.85	&	46101.91	&	-	&	-	&	 - 	&	Dusty	\\
NGC4565	&	NICMOS	&	F160W	&	16	&	1.44	&	3.9	&	18.3	&	61.7	&	18.66	&	Dusty	\\
NGC4579	&	WFPC2	&	F791W	&	19.24	&	3.32	&	9.77	&	20.22	&	48.67	&	19.39	&	Dusty	\\
NGC4772	&	WFPC2	&	F606W	&	23.7	&	6.27	&	68.5	&	21.68	&	36.77	&	 - 	&	Dusty	\\
NGC4826	&	NICMOS	&	F160W	&	16.68	&	2.21	&	6.87	&	19.5	&	80.47	&	 - 	&	Dusty	\\
NGC4866	&	ACS	&	F814W	&	33.20	&	13.65	&	33227.88	&	-	&	-	&	 - 	&	Dusty	\\
NGC5448	&	NICMOS	&	F160W	&	17.36	&	1.49	&	2.56	&	20.71	&	130.06	&	 - 	&	Dusty	\\
NGC5566	&	WFPC2	&	F606W	&	17.16	&	0.86	&	5.23	&	-	&	-	&	 - 	&	Dusty	\\
NGC5866	&	NICMOS	&	F160W	&	18.26	&	1.77	&	23.21	&	-	&	-	&	 - 	&	Dusty	\\
NGC5985	&	NICMOS	&	F160W	&	19.56	&	2.03	&	4.91	&	-	&	-	&	22.46	&	Complex	\\
NGC6340	&	ACS	&	F814W	&	20.28	&	3.44	&	5.07	&	22.14	&	27.77	&	 - 	&	Dusty	\\
NGC7177	&	NICMOS	&	F160W	&	18.16	&	1.89	&	5.97	&	-	&	-	&	 - 	&	Dusty	\\
NGC7331	&	NICMOS	&	F160W	&	22.96	&	8.09	&	344.32	&	18	&	21.64	&	 - 	&	Dusty	\\
NGC7814	&	NICMOS	&	F160W	&	17.82	&	1.91	&	13.16	&	-	&	-	&	 - 	&	Dusty	\\
\hline
\multicolumn{11}{c}{Double S\'{e}rsic Galaxies} \\
Galaxy & Instrument & Filter & $\mu_{e1}^{a}$ & $n_{1}$ & $r_{e1}^{b}$ & $\mu_{e2}^{a}$ & $n_{2}$ & $r_{e2}^{b}$ & \hspace{-0.1in} $\mu_{e, disc}$ & \hspace{-0.1in} $r_{e, disc}$ & \hspace{-0.1in} $m_{pt}^{c}$ & Uncertainty \\
\hline
NGC2768	&	WFPC2	&	F814W	&	17.14	&	1.00	&	0.44	&	24.88	&	5.98	&	547.44	&	-	&	-	&	 - 	&	Dusty	\\
NGC2787	&	WFPC2	&	F814W	&	16.74	&	1.54	&	0.58	&	19.78	&	2.20	&	20.34	&	-	&	-	&	 - 	&	Dusty	\\
NGC2859	&	ACS	&	F814W	&	20.00	&	0.16	&	0.31	&	20.20	&	2.32	&	9.46	&	-	&	-	&	-	&	Complex	\\
NGC3675	&	NICMOS	&	F160W	&	19.14	&	0.17	&	8.96	&	21.11	&	5.51	&	42.71	&	21.11	&	1141.66	&	 - 	&	Dusty	\\
NGC3705	&	WFPC2	&	F814W	&	14.32	&	0.4	&	0.22	&	20.58	&	4.01	&	28.64	&	20.82	&	215.19	&	 - 	&	Dusty	\\
NGC4111	&	NICMOS	&	F160W	&	17.58	&	0.16	&	1.41	&	18.75	&	3.69	&	19.13	&	-	&	-	&	 - 	&	Complex	\\
NGC4220	&	ACS	&	F814W	&	20.60	&	2.25	&	5.57	&	20.29	&	0.14	&	46.12	&	-	&	-	&	 - 	&	Dusty	\\
NGC4293	&	NICMOS	&	F160W	&	16.4	&	0.43	&	0.82	&	29.77	&	8.65	&	25512.11	&	22.23	&	282.46	&	17.99	&	Dusty	\\
NGC4394	&	ACS	&	F814W	&	19.83	&	0.16	&	1.02	&	21.45	&	3.81	&	9.06	&	22.03	&	24.62	&	 - 	&	Dusty	\\
NGC4501	&	WFPC2	&	F606W	&	15.97	&	0.73	&	0.64	&	19.06	&	1.11	&	12.09	&	22.08	&	202.65	&	19.89	&	Dusty	\\
NGC4569	&	NICMOS	&	F160W	&	15.15	&	3.72	&	0.72	&	15.86	&	0.59	&	2.36	&	18.95	&	22.24	&	18	&	Dusty	\\
NGC4596	&	WFPC2	&	F606W	&	15.54	&	0.89	&	0.07	&	22.45	&	4.22	&	66.19	&	-	&	-	&	 - 	&	Dusty	\\
NGC4750	&	NICMOS	&	F160W	&	14.46	&	0.16	&	0.07	&	17.6	&	1.36	&	2.39	&	20.36	&	157.09	&	 - 	&	Dusty	\\
\end{tabular}
\end{minipage}
\end{table*}

We included an exponential disc component in the models for the spiral and lenticular galaxies in our sample, however in many cases this component was negligible. We therefore only include the disc component parameters in tables~\ref{paramconf} and \ref{paramuncert} for the galaxies where it is significant.

To enable us to compare our results with previous studies such as \citet{cb}, which used the Nuker classification scheme, we also fitted the Nuker model to all galaxies in our sample. If we assume that `intermediate' as well as `core' galaxies from the Nuker scheme correspond to `Core-S\'{e}rsic' galaxies and `Double-S\'{e}rsic' and `S\'{e}rsic' galaxies both correspond to `power-law' galaxies in the Nuker scheme then the Nuker classifications of the galaxies in our sample agree with those that we present in table~\ref{paramconf} for 82\% of the galaxies for which we have a confident fit. This suggests that while the classification of many galaxies in our sample remain unchanged when we use the Nuker model, there are still some discrepancies between the two classification schemes.

\section{Discussion}
\begin{figure*}
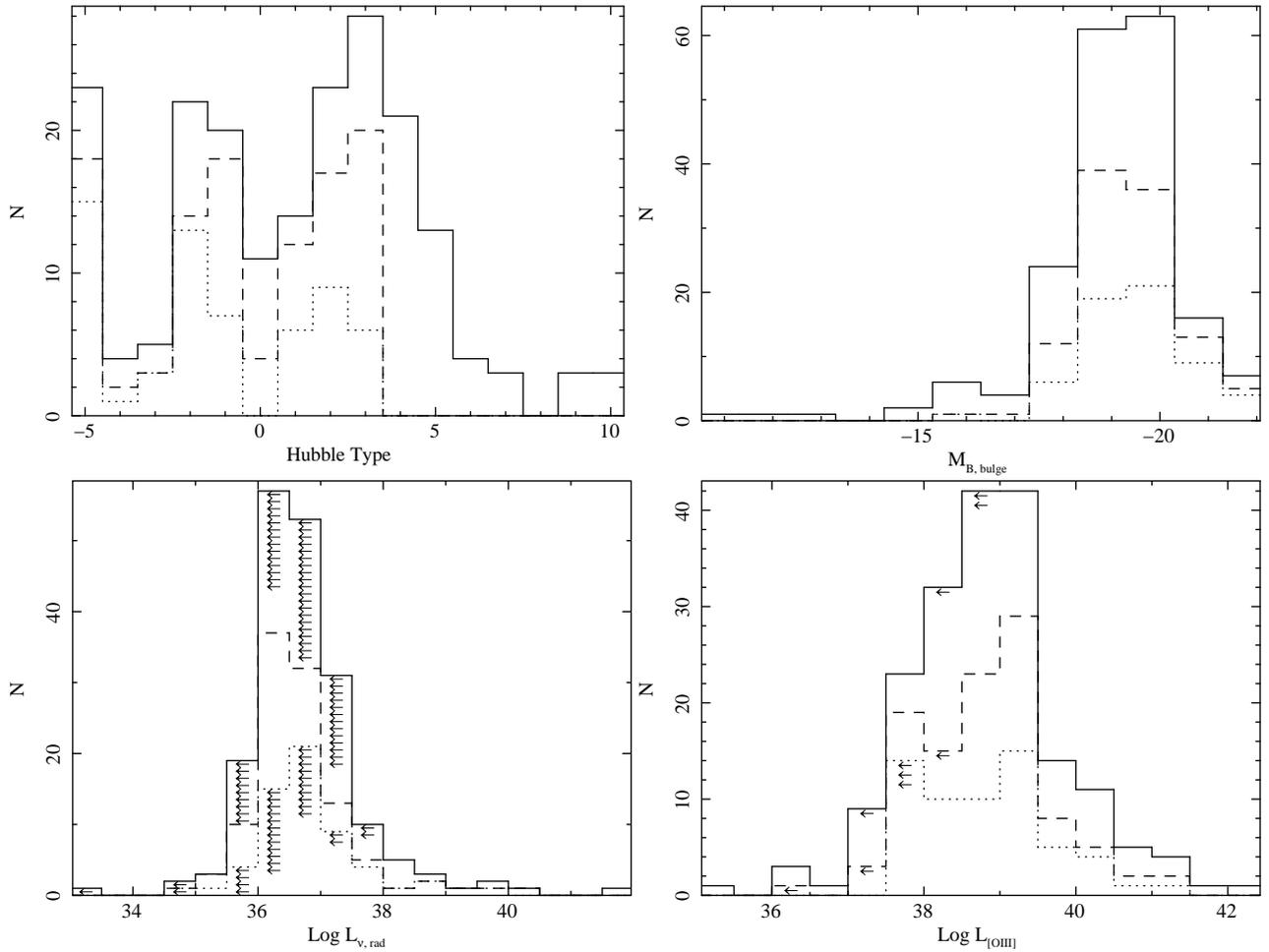

\mbox{\includegraphics[angle=270, width=84mm]{fig3a.ps}}
\mbox{\includegraphics[angle=270, width=84mm]{fig3b.ps}}
\mbox{\includegraphics[angle=270, width=84mm]{fig3c.ps}}
\mbox{\includegraphics[angle=270, width=84mm]{fig3d.ps}}
\caption{Histograms showing the distributions of Hubble type (top left), bulge B band magnitude (top right), nuclear radio luminosity (bottom left) and [O{\rm {\sc iii}}] line luminosity (bottom right) in our final sample (dotted line), the galaxies for which we have a confident or uncertain fit and luminosity data (dashed line) and the Nagar sample (solid line). The arrows denote numbers of galaxies in the histogram bins which are limits only.}
\label{biascomp}
\end{figure*}

Of the 62 galaxies for which we are confident in the brightness profile fits, we have 21 Core galaxies, 26 S\'{e}rsic galaxies and 15 Double-S\'{e}rsic galaxies. We only have both luminosity data and a confident classification for 60 of the 197 galaxies in the Palomar sample that are confirmed as AGN hosts, so we need to verify that we have not introduced any observational biases, for example if bright galaxies are more likely to be imaged with {\it HST} then there would be a disproportionately large number of them in our final sample. In fig.~\ref{biascomp} we compare the distributions of the Hubble type, bulge B band magnitude, nuclear radio luminosity and [O{\rm {\sc iii}}] line luminosity for the 60 galaxies in our final sample, the 107 galaxies for which we have a confident or uncertain fit and luminosity data, and the 197 galaxies hosting AGN in the Palomar sample, that were studied in \citet{na} (the Nagar sample). We used the ASURV Rev 1.2 software \citep*{lavalley92} to test the significance of correlations and differences between distributions for censored data, which includes non-detections with limits. This software implements the methods presented in \citet{feigelson85} for univariate problems and \citet*{isobe86} for bivariate problems. 

According to the generalised Wilcoxon tests the distributions of Hubble type in our final sample and the Nagar sample are inconsistent with being drawn from the same distribution at the 99.99 per cent confidence level. We can see in fig.~\ref{biascomp}, top left panel, that our final sample has a higher proportion of elliptical galaxies, which is to be expected because we discarded all late spirals, and a disproportionately high number of the fits for early spirals were uncertain because they are more likely to contain dust. Using the generalised Wilcoxon tests, we also find that the distributions of bulge magnitude in our final sample and the Nagar sample are only marginally inconsistent with being drawn from the same distribution at the 90 per cent confidence level. There are a small number of galaxies in the Nagar sample with very dim bulges that we missed because we discarded late type spirals. Also, we were more likely to find {\it HST} images for brighter galaxies. Of the galaxies for which we did obtain {\it HST} data, a high proportion of those with intermediate bulge magnitudes had uncertain Nuker fits, probably because these tend to correspond to later type galaxies which are more likely to be dusty. According to the generalised Wilcoxon tests the distributions of nuclear radio luminosities in our final sample and the Nagar sample are also marginally inconsistent with being drawn from the same distribution at the 90 per cent confidence level. From fig.~\ref{biascomp}, bottom left panel,  we can see that there is a higher proportion of the small number of AGNs with higher radio luminosities in our final sample. Finally, using the generalised Wilcoxon tests we find that the distributions of [O{\rm {\sc iii}}] line luminosities in our final sample and the Nagar sample are consistent with being drawn from the same distribution, although we can see in fig.~\ref{biascomp}, bottom right panel, that the galaxies with the lowest [O{\rm {\sc iii}}] line luminosities do not appear in our final sample.

\subsection{Double-S\'{e}rsic Galaxies}

We find 15 galaxies that are best fit by a Double-S\'{e}rsic model. Previous studies \citep[e.g.][]{cote07} attributed the inner S\'{e}rsic component in such models to a resolved stellar nucleus, with a radius $\approx 0.02 r_{e}$, where $r_{e}$ is the effective radius of the outer component. However, several of the Double-S\'{e}rsic galaxies in our sample have inner components with a much larger radius. This suggests that they might not be due to resolved stellar nuclei, for example some of them may instead be inner discs or nuclear bars \citep{erwin}. We studied the isophotes of the galaxies and we did not find any with flattened isophotes towards the centre, which suggests that nuclear bars are unlikely. However, searching through the literature we find that inner discs have been discovered in some of the galaxies in our sample, for example NGC 3945 \citep{moiseev} and NGC 7217 \citep{zasov}. We therefore cannot be certain whether the Double-S\'{e}rsic galaxies in our sample contain resolved stellar nuclei, or if they instead contain other nuclear structures such as inner discs.

\subsection{Black hole mass and bulge magnitude}
\label{basicprop2}

\begin{figure*}
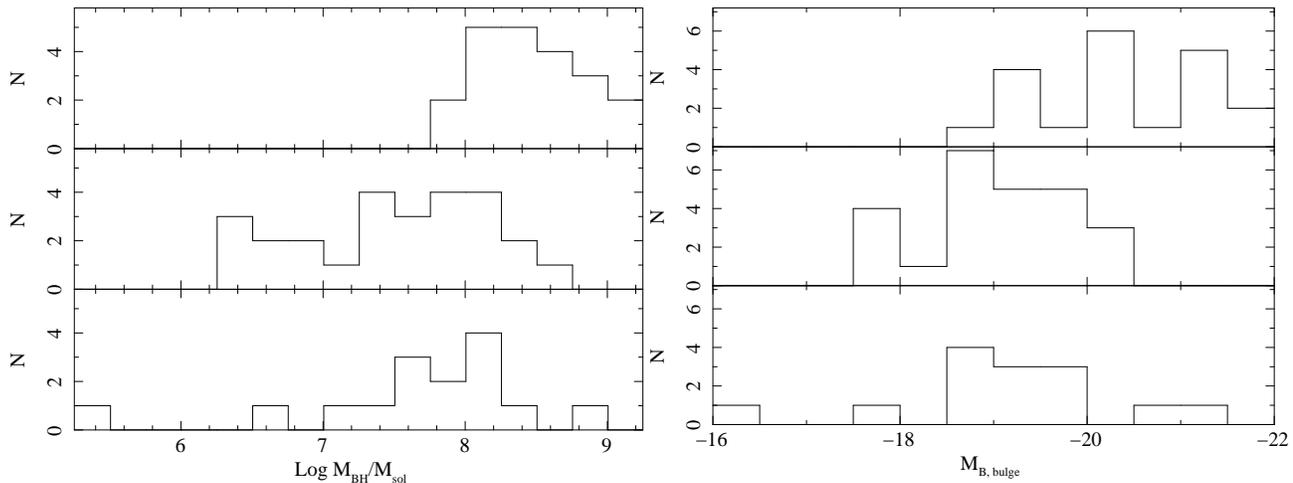

\mbox{
	\includegraphics[angle=270, width=84mm]{fig4a.ps}
	\includegraphics[angle=270, width=87mm]{fig4b.ps}}
\caption{Histograms showing the distributions of central black hole mass (left) and bulge B band magnitude (right) for Core galaxies (top), S\'{e}rsic galaxies (middle) and Double-S\'{e}rsic galaxies (bottom).}
\label{mbh-mag}
\end{figure*}

Fig.~\ref{mbh-mag}, left panel, shows the distributions of central black hole mass for Core, S\'{e}rsic and Double-S\'{e}rsic galaxies. The black hole masses of the Core galaxies are typically higher than those of the S\'{e}rsic or Double-S\'{e}rsic galaxies, for example the median of $\log M_{BH}/M_{\odot}$ is 8.39 for the Core galaxies, compared to 7.57 and 7.84 for the S\'{e}rsic and Double-S\'{e}rsic galaxies respectively. We also find that the most massive black holes are almost exclusively found in Core galaxies --- with the exception of one Double-S\'{e}rsic galaxy, NGC 3998, all SMBHs with $\log M_{BH}/M_{\odot} \ga 8.5$ are found in Core galaxies. Conversely, all SMBHs with $\log M_{BH}/M_{\odot} \la 7.6$ are found in S\'{e}rsic or Double-S\'{e}rsic galaxies. This is consistent with previous studies \citep[e.g.][]{ca}. 

Fig.~\ref{mbh-mag}, right panel, shows the distributions of bulge B band magnitude for Core, S\'{e}rsic and Double-S\'{e}rsic galaxies. The bulges of Core galaxies are generally brighter than those of S\'{e}rsic or Double-S\'{e}rsic galaxies, although they overlap significantly --- for all Core galaxies $M_{B, bulge} \la -18.6$, and for all S\'{e}rsic galaxies $M_{B, bulge} \ga -20.4$. Most Double-S\'{e}rsic galaxies are within the same range as the S\'{e}rsic galaxies, although there are two examples that are more luminous and one that is much less luminous. \citet{cb} found a similar trend in K band magnitudes, although they also found core galaxies covering a much broader range of magnitudes. It is also worth noting that, with the exception of one Double-S\'{e}rsic galaxy, NGC 4125, all galaxies with bulge B magnitudes below -21 are Core galaxies.

We note that the most massive black holes being found in Core galaxies is to be expected as they are at the more luminous end of the luminosity function \citep[e.g.][and also confirmed for our sample by Figure~\ref{mbh-mag}, right panel]{ferrarese06b}, so they also tend to have larger velocity dispersions \citep{faber} and hence black hole masses, which we derived from the velocity dispersions.

\subsection{Radio-Loudness}

The radio-loudness of an AGN is typically measured as the ratio of the radio to optical luminosity of the nucleus, however we were only able to extract a point source in 23 of the galaxies for which we are confident in the Core/S\'{e}rsic fits, so this would limit the number of nuclei for which we can measure the radio-loudness. Furthermore this could bias any correlations that we observe because we would be preferentially selecting bright, type 1 nuclei that are unobscured by nuclear dust. We therefore use the narrow [O{\rm {\sc iii}}] line luminosity instead of the total optical luminosity of the nucleus, because we have this data for most of the galaxies in our sample, and since [O{\rm {\sc iii}}] line emission originates from the narrow line region, there should not be a bias between type 1 and type 2 nuclei.

\begin{figure}
\mbox{\includegraphics[angle=270, width=84mm]{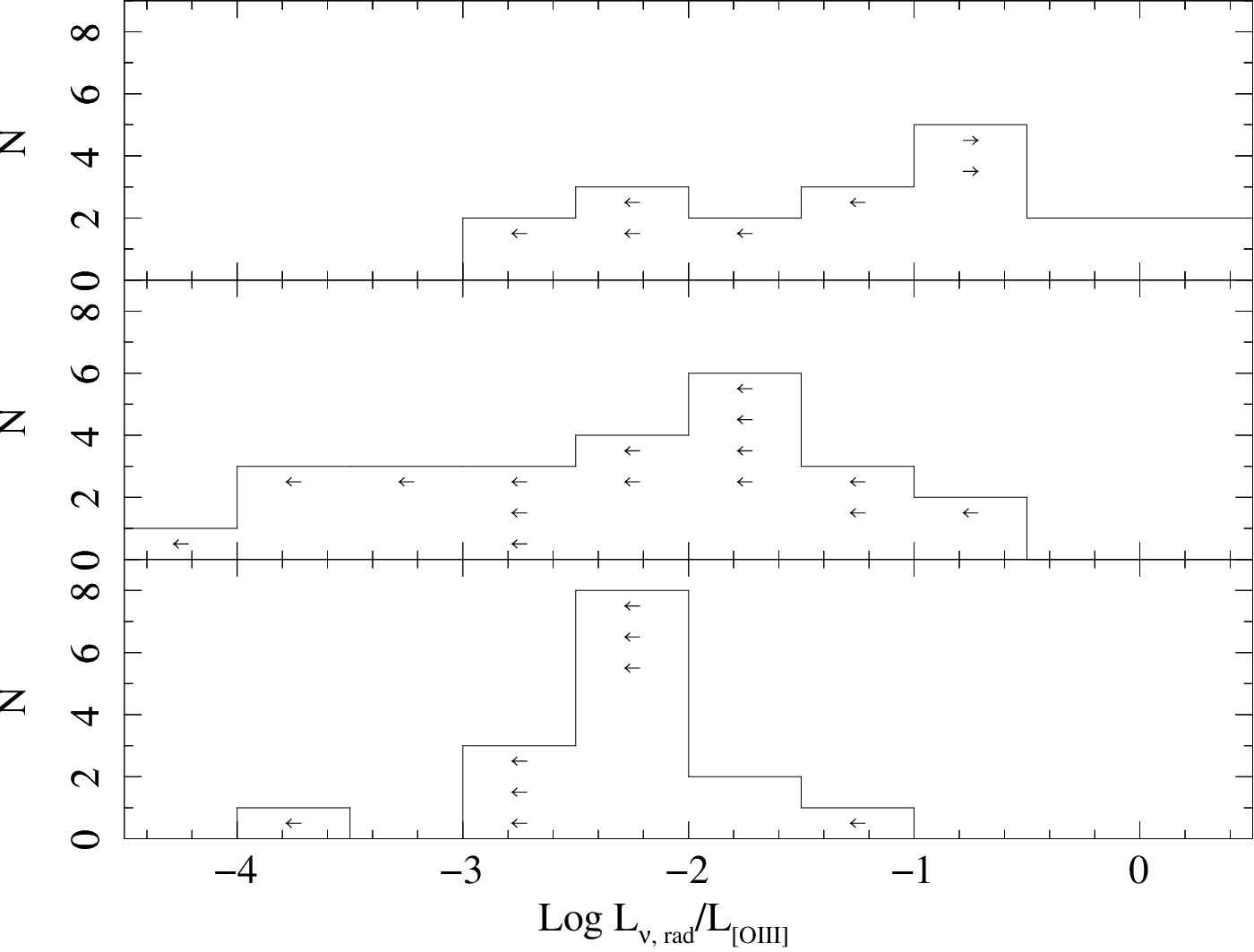}}
\caption{Histograms showing the distribution of $L_{\nu, rad}/L_{[O{\rm III}]}$ for Core galaxies (top), S\'{e}rsic galaxies (middle) and Double-S\'{e}rsic galaxies (bottom).  The arrows denote numbers of galaxies in the histogram bins which are limits only.}
\label{o3radfig}
\end{figure}

\begin{figure*}
\mbox{\includegraphics[angle=270, width=84mm]{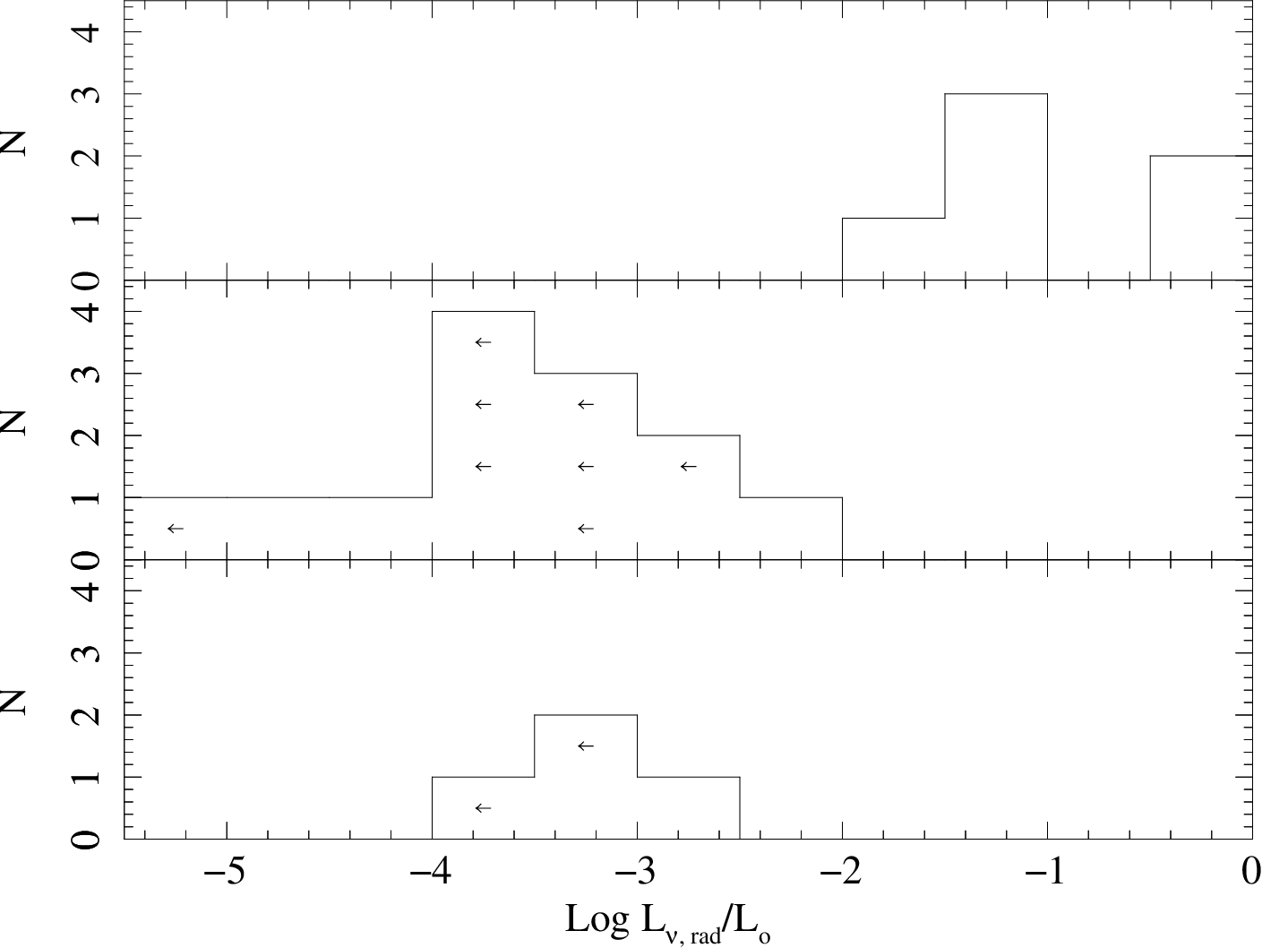}}
\mbox{\includegraphics[angle=270, width=84mm]{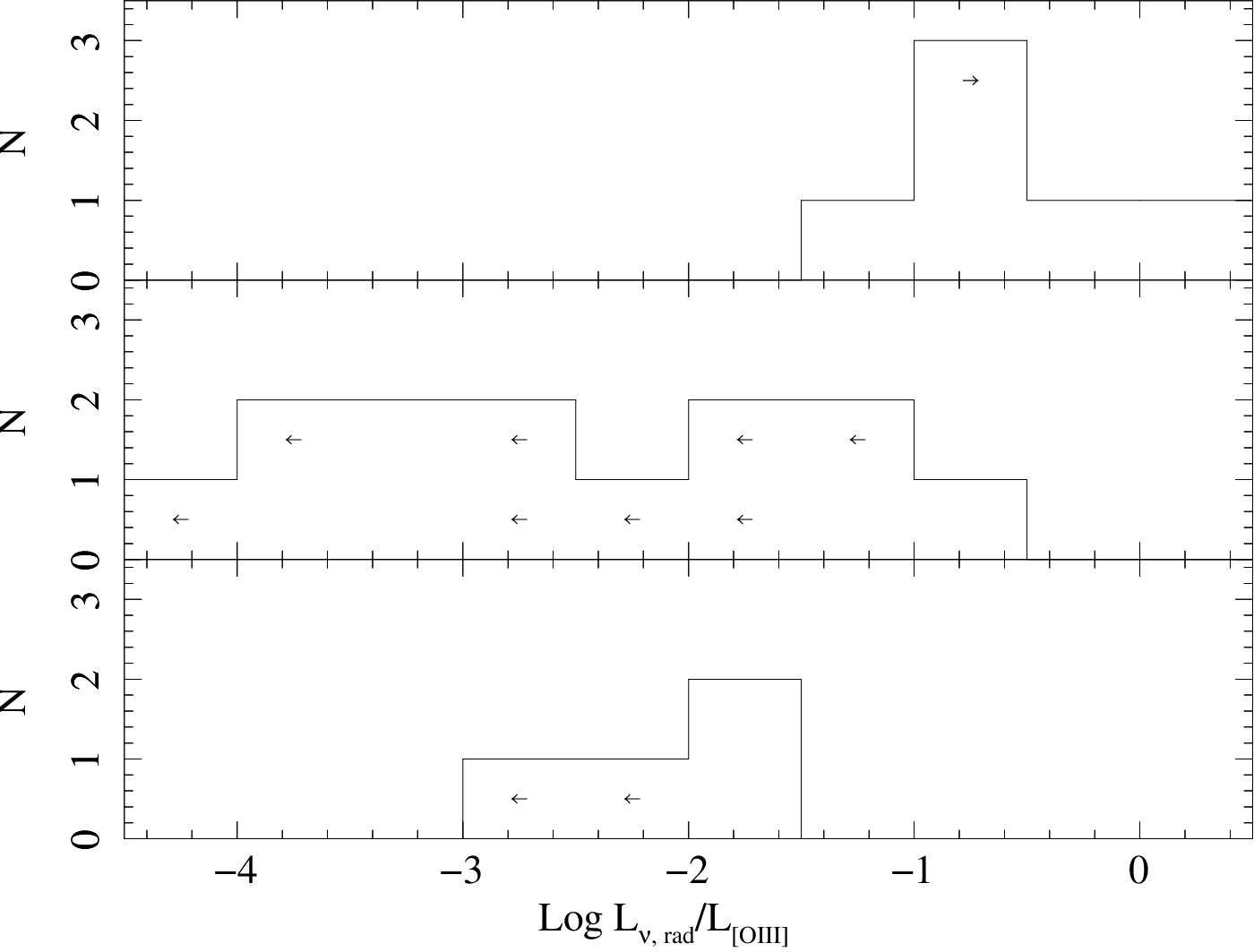}}
\caption{Histograms comparing the distributions of $L_{\nu, radio}/L_o$ (left panel) and $L_{\nu, radio}/L_{[O{\rm III}]}$ (right panel) for the Core (top), S\'{e}rsic (middle) and Double-S\'{e}rsic (bottom) galaxies for which we were able to extract a point source.  The arrows denote numbers of galaxies in the histogram bins which are limits only.}
\label{optradfig}
\end{figure*}

In Fig.~\ref{o3radfig} we show the distribution of the radio-loudness $L_{\nu, radio}/L_{[O{\rm III}]}$ for Core, S\'{e}rsic and Double-S\'{e}rsic galaxies. According to the generalised Wilcoxon tests the distributions of radio-loudness in Core and S\'{e}rsic galaxies are inconsistent with being drawn from the same distribution at the 99.9 per cent confidence level, while the distributions of radio-loudness in S\'{e}rsic and Double-S\'{e}rsic galaxies are consistent with being drawn from the same distribution. This confirms that the radio-loudness of an AGN is related to whether its host galaxy is classified as a Core galaxy. We can see in fig.~\ref{o3radfig} that the most radio-loud AGNs are hosted in Core galaxies, however AGNs with $\log(L_{\nu, radio}/L_{[O{\rm III}]})\la-0.8$ can be found in both Core and S\'{e}rsic galaxies. This large overlap has not been observed in previous studies, for example \citet{ca} found that almost all core galaxies were more radio-loud than the power-law galaxies (classified using the Nuker scheme), with very little overlap. Their sample was restricted to early-type galaxies whereas our sample also included early spiral galaxies, however excluding the spirals in our sample did not recover the obvious split that they observed. The optical nuclear luminosities reported in \citet{ca} are higher for power-law galaxies than for core galaxies, however the lowest [O{\rm {\sc iii}}] line luminosities of power-law galaxies are as low as those of core galaxies, although they do extend to higher [O{\rm {\sc iii}}] line luminosities than core galaxies. This suggests that the discrepancy between the correlation that we observe and that observed by \citet{ca} is due to the different definitions used for the radio-loudness. To test this possibility we looked at the galaxies for which we were able to extract a point source in the profile fitting and we used this point source to determine the ratio of nuclear radio luminosity to nuclear optical continuum luminosity, $L_{\nu, radio}/L_o$. For these galaxies we then compared the distributions of both $L_{\nu, radio}/L_o$ and $L_{\nu, radio}/L_{[O{\rm III}]}$ for Core, S\'{e}rsic and Double-S\'{e}rsic galaxies; these results are presented in Fig.~\ref{optradfig}. According to the generalised Wilcoxon tests the distributions of $L_{\nu, radio}/L_o$ in Core and S\'{e}rsic galaxies are inconsistent with being drawn from the same distribution at the 99.99 per cent confidence level, while the distributions of $L_{\nu, radio}/L_{[O{\rm III}]}$ are inconsistent at the 99.9 per cent level. Furthermore, we can see from Fig.~\ref{optradfig} that there is no overlap between the Core and S\'{e}rsic subsamples when we use the nuclear optical continuum luminosity. This demonstrates that the split between Core and S\'{e}rsic galaxies is clearer when we use $L_{\nu, radio}/L_o$.

\begin{figure}
\includegraphics[angle=270, width=84mm]{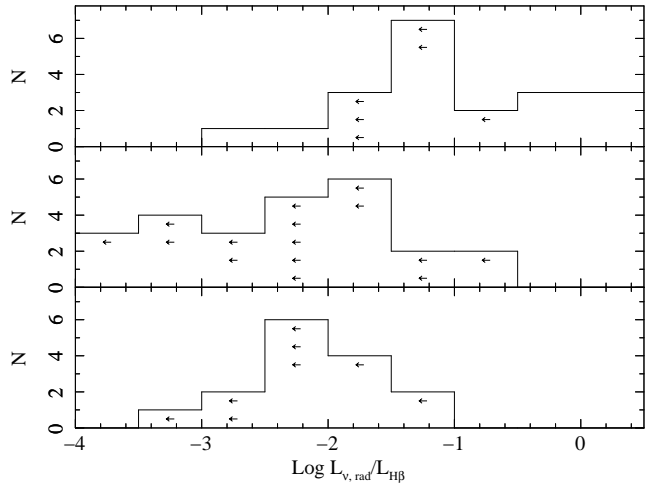}
\caption{Histograms showing the distribution of $L_{\nu, rad}/L_{H\beta}$ for Core (top), S\'{e}rsic (middle) and Double-S\'{e}rsic (bottom) galaxies. The arrows denote numbers of galaxies in the histogram bins which are limits only.}
\label{hbradhist}
\end{figure}

We also looked at using the narrow H$\beta$ line luminosity instead of [O{\rm {\sc iii}}]. Fig.~\ref{hbradhist} shows that using this definition for the radio-loudness produces the same difference between Core and S\'{e}rsic galaxy radio-loudness distributions that was observed when [O{\rm {\sc iii}}] was used.

\subsection{Correlations of radio-loudness with black hole mass and bulge magnitude}

\begin{figure*}
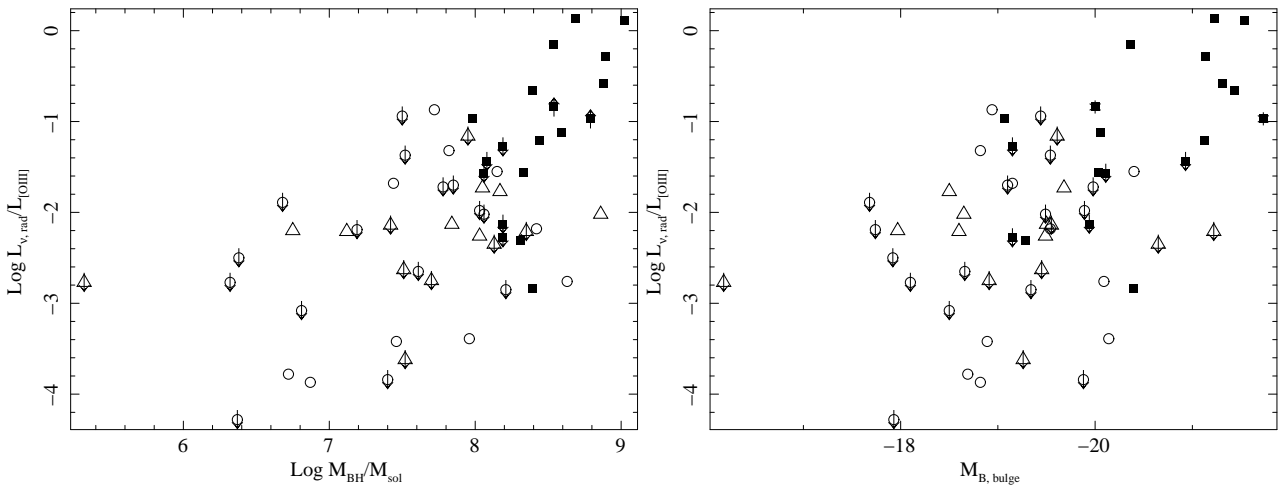

\mbox{\includegraphics[angle=270, width=84mm]{fig8a.ps}}
\mbox{\includegraphics[angle=270, width=84mm]{fig8b.ps}}
\caption{Radio-loudness $L_{\nu, rad}/L_{[O{\rm III}]}$ plotted against black hole mass (left panel) and against bulge B band magnitude (right panel). Core galaxies are represented by filled squares, S\'{e}rsic galaxies by empty circles and Double-S\'{e}rsic galaxies by empty triangles.}
\label{radvothfig}
\end{figure*}

We saw in \S~\ref{basicprop2} that the distributions of both the black hole mass and the bulge B band magnitude are different for Core and S\'{e}rsic galaxies, so this could lead to the observed trends of radio-loudness with the Core/S\'{e}rsic classification if radio-loudness is dependent on one of these properties. To test this possibility we plot radio-loudness against the black hole mass and the bulge B band magnitude; these plots are presented in Fig.~\ref{radvothfig}.

Using Spearman's rho to test the correlation between radio-loudness and black hole mass, we find that there is a correlation at the 99.99 per cent confidence level, with galaxies hosting high mass black holes generally being more radio-loud. Similar correlations have been demonstrated by previous studies of AGN, for example radio-loud quasars are found to contain systematically more massive black holes than radio-quiet quasars \citep{laor, mclure}, and \citet{doi} found a best-fit plane in the three dimensional space of radio-loudness, black hole mass and Eddington ratio for a sample of 48 LLAGN taken from the Palomar spectroscopic survey. In fig.~\ref{radvothfig}, left panel, we can see that there is still considerable scatter between these variables, which suggests that other factors may be involved. 

We also used Spearman's rho to test the correlation between radio-loudness and bulge B band magnitude, and we found that there is a correlation at the 99 per cent confidence level, with luminous galaxies generally being more radio loud. However, fig.~\ref{radvothfig}, right panel shows that there is also scatter between these two variables, which again suggests that there may be other factors involved.

It is possible that the scatter in the correlations of radio-loudness with bulge magnitude and black hole mass can be explained by the Core/S\'{e}rsic/Double-S\'{e}rsic classification. To test correlations with this classification we parameterised the galaxies using the central deficit or excess in the surface brightness profile compared to the inward extrapolation of the outer S\'{e}rsic component, using the $\Delta_{0.02}$ parameter introduced by \citet{cote07}. This parameter is defined as $\Delta_{0.02} = \log \cal{L}_G/\cal{L}_{S}$, where $\cal{L}_G$ is the total luminosity of the best fit galaxy model within a radius of $0.02 r_{e}$ and $\cal{L}_{S}$ is the total luminosity of the outer S\'{e}rsic component in this region. Core galaxies have $\Delta_{0.02} < 0$, S\'{e}rsic galaxies have $\Delta_{0.02} = 0$ and Double-S\'{e}rsic galaxies have $\Delta_{0.02} > 0$. We then tested the partial correlations for both black hole mass and bulge magnitude with radio loudness and $\Delta_{0.02}$, using the cens\_tau code \footnote{http://astrostatistics.psu.edu/statcodes/cens\_tau}, which uses the methods presented in \citet{akritas96} to search for partial correlations in censored data. These partial correlations test whether there are residual correlations between radio-loudness and $\Delta_{0.02}$ after the overall correlations have been accounted for. 

\begin{figure}
\includegraphics[angle=270, width=84mm]{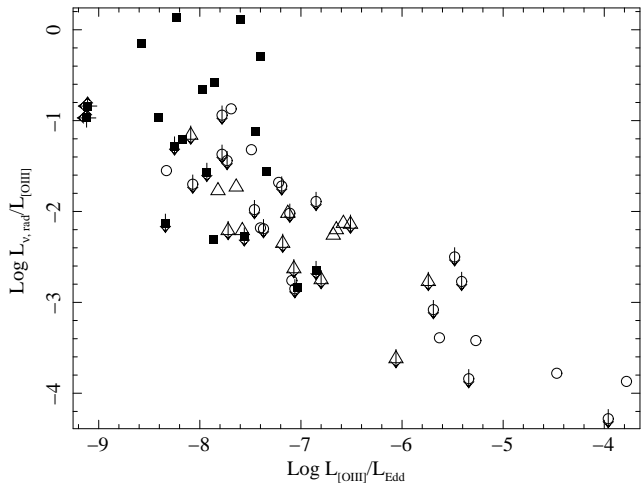}
\caption{Radio-loudness $L_{\nu, rad}/L_{[OIII]}$ plotted against the [O{\rm {\sc iii}}] line luminosity as a fraction of the Eddington luminosity, $L_{[OIII]}/L_{Edd}$, for Core galaxies (filled squares), S\'{e}rsic galaxies (empty circles) and Double-S\'{e}rsic galaxies (empty triangles).}
\label{acc-rate}
\end{figure}

Zero partial correlation between the bulge magnitude, radio-loudness and $\Delta_{0.02}$ is rejected at the 99 per cent confidence level, and zero partial correlation between the black hole mass, radio-loudness and $\Delta_{0.02}$ is rejected at the 99.99 per cent confidence level. These results suggest that there is a significant residual correlation between the radio-loudness of an AGN and the classification of its host galaxy as Core, S\'{e}rsic or Double-S\'{e}rsic, after the effects of bulge magnitude and black hole mass have been accounted for. However, \citet{doi} demonstrated that the radio-loudness correlates with both the black hole mass and the accretion rate in a sample of 48 LLAGN, so we also need to consider whether $\Delta_{0.02}$ correlates with the accretion rate. We used the ratio of [O{\rm {\sc iii}}] line luminosity to Eddington luminosity, $\log(L_{[OIII]}/L_{Edd})$, as a proxy for the accretion rate and tested the partial correlation of $\Delta_{0.02}$ with $\log(L_{[OIII]}/L_{Edd})$ and black hole mass, following the same method as above. We found that $\Delta_{0.02}$ is partially correlated with $\log(L_{[OIII]}/L_{Edd})$ at the 99.99 per cent confidence level. Therefore it is possible that the partial correlation between radio-loudness and $\Delta_{0.02}$ arises because Core galaxies tend to have a higher central black hole mass and a lower accretion rate. This possibility needs further investigation. The correlation between accretion rate and the classification of the host galaxy as Core, S\'{e}rsic or Double-S\'{e}rsic is illustrated in fig.~\ref{acc-rate}, where we plot $\log L_{\nu, rad}/L_{[OIII]}$ against $\log L_{[OIII]}/L_{Edd}$ for the Core, S\'{e}rsic and Double-S\'{e}rsic galaxies.

\section{Conclusions}
In this paper we have studied the brightness profiles of elliptical and early type spiral galaxies in the Palomar spectroscopic survey that are confirmed as AGN hosts. We fitted S\'{e}rsic, Core-S\'{e}rsic and, where necessary, Double-S\'{e}rsic models, plus a point source where needed, to the 1D semi-major axis brightness profiles extracted from high resolution {\it HST} images of these galaxies. By comparing these fits we were able to classify the galaxies as Core, S\'{e}rsic or Double-S\'{e}rsic galaxies, and then we investigated how other properties of the host galaxy and the AGN relate to this classification.

We found that Core galaxies were generally more luminous and hosted higher mass black holes than S\'{e}rsic and Double-S\'{e}rsic galaxies, although there was considerable overlap between these subsamples. These results agree with previous studies \citep[e.g.][who used the Nuker classification scheme]{ca}.

To measure the radio-loudness of the AGNs we took the ratio of nuclear radio luminosity to [O{\rm {\sc iii}}] line luminosity, because we could only estimate the optical continuum luminosity of the AGN for approximately one third of the galaxies, for which we could extract a point source in the profile fitting.  Furthermore, using the [O{\rm {\sc iii}}] line luminosity prevents bias against faint nuclei that are more difficult to extract. Using this definition of the radio-loudness we found that Core galaxies were generally more radio-loud than S\'{e}rsic and Double-S\'{e}rsic galaxies, in agreement with the results of \citet{ca} for a radio-selected sample of AGN based on the Nuker scheme.  However, we found significantly more overlap between these subsamples compared to \citet{ca}. This difference is most likely because we use a different definition of the radio loudness, as we find that there is much less overlap when we use the nuclear optical continuum luminosity. Therefore we conclude that the radio-loudness/brightness-profile connection uncovered for radio-selected AGN also applies to our optically-selected sample.

We also looked at how the radio-loudness, defined using the [O{\rm {\sc iii}}] line luminosity, correlated with the black hole mass and bulge B band magnitude of the host galaxies.  We found that, while the radio -loudness does show correlations with both black hole mass and bulge magnitude, which could explain at least some of the correlation between brightness-profile and radio-loudness, we still found a significant partial correlation with the classification of the host galaxy brightness profile as Core or S\'{e}rsic. However, the host galaxy classification is also correlated with the accretion rate, so it is possible that the observed connection between brightness-profile and radio-loudness arises because Core galaxies tend to have a higher black hole mass and a lower accretion rate. This possibility requires further investigation.

\section*{Acknowledgments}
AJR acknowledges the support of a Nuffield Foundation Undergraduate Research Bursary. PU is supported by an STFC Advanced Fellowship, and funding from the European Community's Seventh Framework Programme (FP7/2007-2013) under grant agreement number ITN 215212 “Black Hole Universe”. This work is based on observations made with the NASA/ESA Hubble Space Telescope, and obtained from the Hubble Legacy Archive, which is a collaboration between the Space Telescope Science Institute (STScI/NASA), the Space Telescope European Coordinating Facility (ST-ECF/ESA) and the Canadian Astronomy Data Centre (CADC/NRC/CSA). We also thank the referee, Laura Ferrarese, for useful comments that have improved this paper.

\label{lastpage}

\end{document}